\documentclass[
prl,
superscriptaddress,
noshowpacs,
twocolumn,
]
{revtex4-2}
\usepackage{graphicx}
\usepackage[colorlinks,
            linkcolor=blue,
            citecolor=blue,
            anchorcolor=blue,
            urlcolor=blue,
            ]{hyperref}
\usepackage{leftidx}
\usepackage{subfigure}
\usepackage{amsmath}
\usepackage{amssymb}
\usepackage{epstopdf}
\usepackage{amsmath}
\usepackage{lineno}

\newlength{\onecolfig}
\newlength{\twocolfig}
\newcommand{\unit}[1]{\,\mbox{#1}}

\newcommand{\MHz}{\unit{MHz}}
\newcommand{\GHz}{\unit{GHz}}

\newcommand{\mK}{\unit{mK}}

\newcommand{\ns}{\unit{ns}}

\newcommand{\dB}{\unit{dB}}

\newcommand{\tr}{\operatorname{tr}}

\newcommand{\ie}{{\em i.e.}}

\newcommand{\bra}[1]{\mbox{$\left< #1 \right|$}}
\newcommand{\ket}[1]{\mbox{$\left| #1 \right>$}}
\newcommand{\avg}[1]{\mbox{$\langle #1 \rangle$}}

\newcommand{\beqa}{\begin{eqnarray}}
\newcommand{\eeqa}{\end{eqnarray}}
\newcommand{\lp}{\left(}
\newcommand{\rp}{\right)}

\begin{document}
\title{Shortcuts to Adiabaticity for Open Systems in Circuit Quantum Electrodynamics}
\author{Zelong Yin}
\affiliation{Tencent Quantum Laboratory, Tencent, Shenzhen, Guangdong 518057, China}
\author{Chunzhen Li}
\affiliation{Tencent Quantum Laboratory, Tencent, Shenzhen, Guangdong 518057, China}
\author{Jonathan Allcock}
\affiliation{Tencent Quantum Laboratory, Tencent, Shenzhen, Guangdong 518057, China}
\author{Yicong Zheng}
\affiliation{Tencent Quantum Laboratory, Tencent, Shenzhen, Guangdong 518057, China}
\author{Xiu Gu}
\affiliation{Tencent Quantum Laboratory, Tencent, Shenzhen, Guangdong 518057, China}
\author{Maochun Dai}
\affiliation{Tencent Quantum Laboratory, Tencent, Shenzhen, Guangdong 518057, China}
\author{Shengyu Zhang}
\affiliation{Tencent Quantum Laboratory, Tencent, Shenzhen, Guangdong 518057, China}
\author{Shuoming An}
\email{shuomingan@tencent.com}
\affiliation{Tencent Quantum Laboratory, Tencent, Shenzhen, Guangdong 518057, China}
\date{\today}

\begin{abstract}
Shortcuts to adiabaticity (STA) are powerful quantum control methods, allowing quick evolution into target states of otherwise slow adiabatic dynamics. 
Such methods have widespread applications in quantum technologies, and various STA protocols have been demonstrated in closed systems.
However, realizing STA for open quantum systems has presented a greater challenge, due to complex controls required in existing proposals. 
Here we present the first experimental demonstration of STA for open quantum systems, using a superconducting circuit QED system consisting of two coupled bosonic oscillators and a transmon qubit. 
By applying a counterdiabatic driving pulse, we reduce the adiabatic evolution time of a single lossy mode from $800\ns$ to $100\ns$. 
In addition, we propose and implement an optimal control protocol to achieve fast and qubit-unconditional equilibrium of multiple lossy modes.
Our results pave the way for accelerating dynamics of open quantum systems and have potential applications in designing fast open-system protocols of physical and interdisciplinary interest, such as accelerating bioengineering and chemical reaction dynamics.
\end{abstract}
\maketitle
\section{Introduction} 
Adiabatic processes -- which preserve the non-degenerate instantaneous eigenstates of time-dependent Hamiltonians -- have important applications in quantum technologies, including quantum simulation~\cite{farhi2001quantum,aharonov2008adiabatic} and computation~\cite{zanardi1999holonomic,ekert2000geometric}.
Though adiabatic evolution is, in principle, relatively robust against parameter fluctuations and environmental  noise~\cite{childs2001robustness}, in the noisy intermediate-scale quantum era, decoherence is still an obstacle preventing its widespread application.
Shortcut to adiabaticity (STA) addresses this issue by finding fast trajectories that connect the initial and final states of slow-paced adiabatic protocols.
Since STA was first proposed~\cite{chen2010fast}, it has found many applications, including in atom cooling, trapped atom and ion transportation~\cite{guery2014transport,an2016shortcuts}, spin population transfer~\cite{bason2012high,zhang2013experimental,du2016experimental, zhou2017accelerated}, implementing quantum logic gates~\cite{martinis2014fast,theis2018counteracting}, and quantum thermodynamics~\cite{an2015experimental}.
Due to freedom in choosing intermediate trajectories, time-dependent control parameters of a system can be adjusted in different ways, resulting in various STA protocols~\cite{guery2019shortcuts}.
In particular, the method of counterdiabatic (CD) driving~\cite{demirplak2003adiabatic,berry2009transitionless} adds an auxiliary control $H_{CD}$ to the reference Hamiltonian to suppress unwanted diabatic transitions. 
This adiabatic-following feature makes CD driving robust to parameter errors~\cite{an2016shortcuts} and suitable for fast holonomic gates~\cite{song2016shortcuts} and efficient quantum heat engines~\cite{beau2016scaling}.

While STA finds widespread application in closed quantum systems, its generalization to open quantum systems is of fundamental interest.
There are two strategies for this generalization: 
first, one can stick with STA designed for closed systems and mitigate environmental effects by utilizing redundant degrees of freedom \cite{zhou2017accelerated};
second, one can directly attempt to accelerate open system adiabatic dynamics.
For open classical systems, a swift-equilibration protocol similar to STA was used to accelerate the equilibration of a Brownian particle \cite{martinez2016engineered}.
Recently, this idea was extended to the field of biology to guide the probability distribution of genotypes in a population along a specified path and time interval \cite{iram2021controlling}.
For open quantum systems, CD driving can be designed theoretically based on non-Hermitian Hamiltonians \cite{ibanez2011shortcuts,chen2016method} or Lindblad dynamics \cite{vacanti2014}.
However, it remains challenging to conduct such experiments, as they often require complex controls such as engineered system-bath interactions \cite{villazon2019swift}.

Here, we generalize STA to an open circuit QED (cQED) system consisting of multiple dissipative bosonic modes. 
When the time-dependent controls are varied sufficiently slowly~\cite{sarandy2005adiabatic}, the system evolves adiabatically within its time-dependent decoherence-free subspace (DFS)~\cite{lidar1998decoherence}.
Analogously to the CD driving for closed quantum systems, we deduce the diabatic part of the Liouvillian which causes non-equilibrium transitions, and add a unitary control to counteract it.
Consequently, we can enforce fast adiabatic evolution of the time-dependent DFS, i.e., a system initialized in the DFS remains so at all times \cite{wu2017adiabatic}.
However, when multiple lossy modes are present -- as is common in many experimental scenarios -- under the one-port driving of our setup (see below), only one hybrid mode at a time can be under perfect CD control.

To realise STA in a multi-mode setting, we further develop an analytical, open-loop control protocol, which we term multi-mode optimal control (MMOC). 
In MMOC, we make use of non-adiabatic dynamics during ringup(ringdown) to achieve the desired final equilibrium in a duration much less than that required by a slow varying adiabatic reference process (Fig.~\ref{main:cd_result}). 
Utilising redundant degrees of freedom, we can also minimise a user-defined merit function, which we choose here to be the maximum driving amplitude to avoid undesired qubit excitations~\cite{sank2016measurement}.
See Fig.~\ref{main:problem_setup}a for a schematic overview of the system dynamics under the CD and MMOC procedures.

\section{Experimental setup}
\begin{figure*}[tbp]
\begin{center}
\includegraphics[width=0.6\twocolfig]{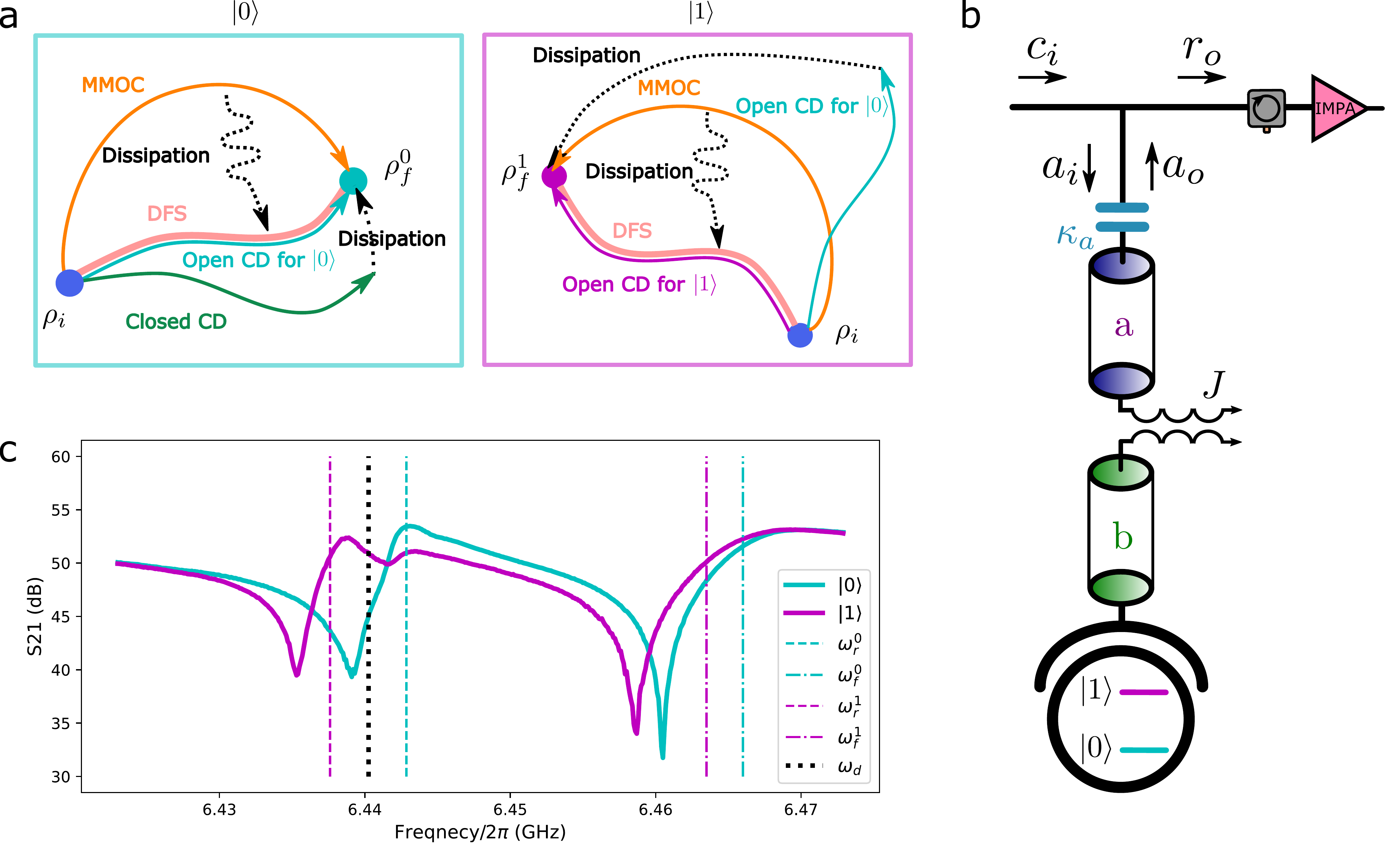}
\end{center}
\caption{\textbf{Experimental setup and principles of shortcut to adiabaticity (STA)} 
(a) Schematic of relevant dynamics. 
An adiabatic protocol transfers the initial state $\rho_i$ to the final state $\rho_f^{0,1}$ along the light pink trajectories in the decoherence-free subspace (DFS), which is precise only in the infinite time limit.
To accelerate this adiabatic trajectory, a counterdiabatic (CD) driving can be applied to cancel non-adiabatic transitions, making the adiabatic approximation exact. Due to the effect of dissipation, CD driving for closed quantum systems has limited effectiveness, while its open system extension achieves arbitrary speed up for a single mode conditioned on the qubit state.
To realize STA for multiple modes and independent of qubit states, we develop the multi-mode optimal control (MMOC) protocol, which includes non-adiabatic trajectories at intermediate times and reaches both final states $\rho_f^{0,1}$ at a given target time.
(b) Setup of the superconducting circuit. 
The resonator mode $b$ is dispersively coupled to a transmon qubit with strength $g/2\pi\approx 80\MHz$. 
The Purcell filter mode $a$ is coupled to the resonator mode with strength $J/2\pi=10.5\MHz$ and to the feedline with strength $\kappa_a/2\pi=11.4\MHz$.
From residue qubit excitation, the temperature of the environment is measured to be $75\mK$, justifying the cold-bath approximation of the feedline.
Driving pulses are applied through the input field $c_i$, and the output field $r_0$ is amplified by an impedance-matched parametric amplifier (IMPA) and homodyne detected to infer the system dynamics.
(c) Measured transmission spectrum $S_{21}$ and hybrid-mode frequencies for both qubit states. 
The driving frequency $\omega_d=2\pi\times 6.44025\GHz$ is allocated roughly in the middle of $\omega_r^0=2\pi\times 6.4427\GHz$ and $\omega_r^1=2\pi\times 6.4378\GHz$, where the $0$ or $1$ in the upper right corner denotes the qubit state.
The significant shift of resonant dips is partly due to the nonlinearity of the cavity when it is driven with large power.
See Supplementary Note 1 for more details on the asymmetry and frequency shifts.
}
\label{main:problem_setup}
\end{figure*}
Our setup is illustrated in Fig.~\ref{main:problem_setup}b and consists of two coupled resonator modes~\cite{zhou2021rapid}, $a$ and $b$, with coupling strength $J/2\pi\approx 10.5 \MHz$.
Mode $b$ is dispersively coupled to a transmon qubit~\cite{koch2007transmon} and mode $a$ is coupled to a feedline at an effective temperature of $75\mK$ with strength $\kappa_a/2\pi\approx 11.4 \MHz$, which serves as a cold bath. 
Given the dispersive coupling to the transmon qubit, the coupled modes $a$ and $b$ can be decoupled into qubit-dependent hybrid modes (normal modes) $a'^{0,1}$ and $b'^{0,1}$ (see Methods). 
A coherent drive with amplitude $\epsilon(t)$ at frequency $\omega_d/2\pi= 6.44025\GHz$ is generated by an arbitrary waveform generator and applied through the feedline. 
The system dynamics is inferred via time-traced output homodyne detection using the input-output theory as detailed in Supplementary Note (SN) 1.

\section{Open System STA by counterdiabatic driving}
We first implement CD driving designed for lossy hybrid mode $b'$ to achieve a fast ringup in a target duration $t_f$, with the qubit kept in the ground state. 
For a chosen reference drive $\epsilon(t)$, the required additional CD control can be derived from the Lindblad dynamics or time-dependent DFS of the system (see SN 2 and 3).
The CD driving in the rotating frame of driving frequency $\omega_d$, including the reference $\epsilon(t)$ and the auxiliary control, takes the form:
\begin{linenomath*}
\begin{equation}
    \epsilon_{CD}(t) = \epsilon(t) -i\frac{\dot{\epsilon}(t)}{\Delta^0_r-i\kappa^0_r/2},
    \label{eq:cd_drive}
\end{equation}
\end{linenomath*}
where $\Delta_r^0 \equiv \omega_r^0-\omega_d$ and $\kappa_r^0$ are the frequency detuning and decay constant respectively, for the hybrid resonator mode $b'^0$ and qubit state in $\left|0\right>$.
For convenience, the superscripts are omitted in what follows. 
Note that in the dissipation-free limit $\kappa_r\to 0$, the closed system CD solution is readily recovered \cite{an2016shortcuts}. 
We characterise the performance of the CD driving in terms of the quantum speed limit for open quantum systems \cite{taddei2013quantum} and conclude that the CD driving has optimal quantum efficiency within the space of all available pulses (see SN 4).

\begin{figure*}[tbp]
\begin{center}
\includegraphics[width=0.55\twocolfig]{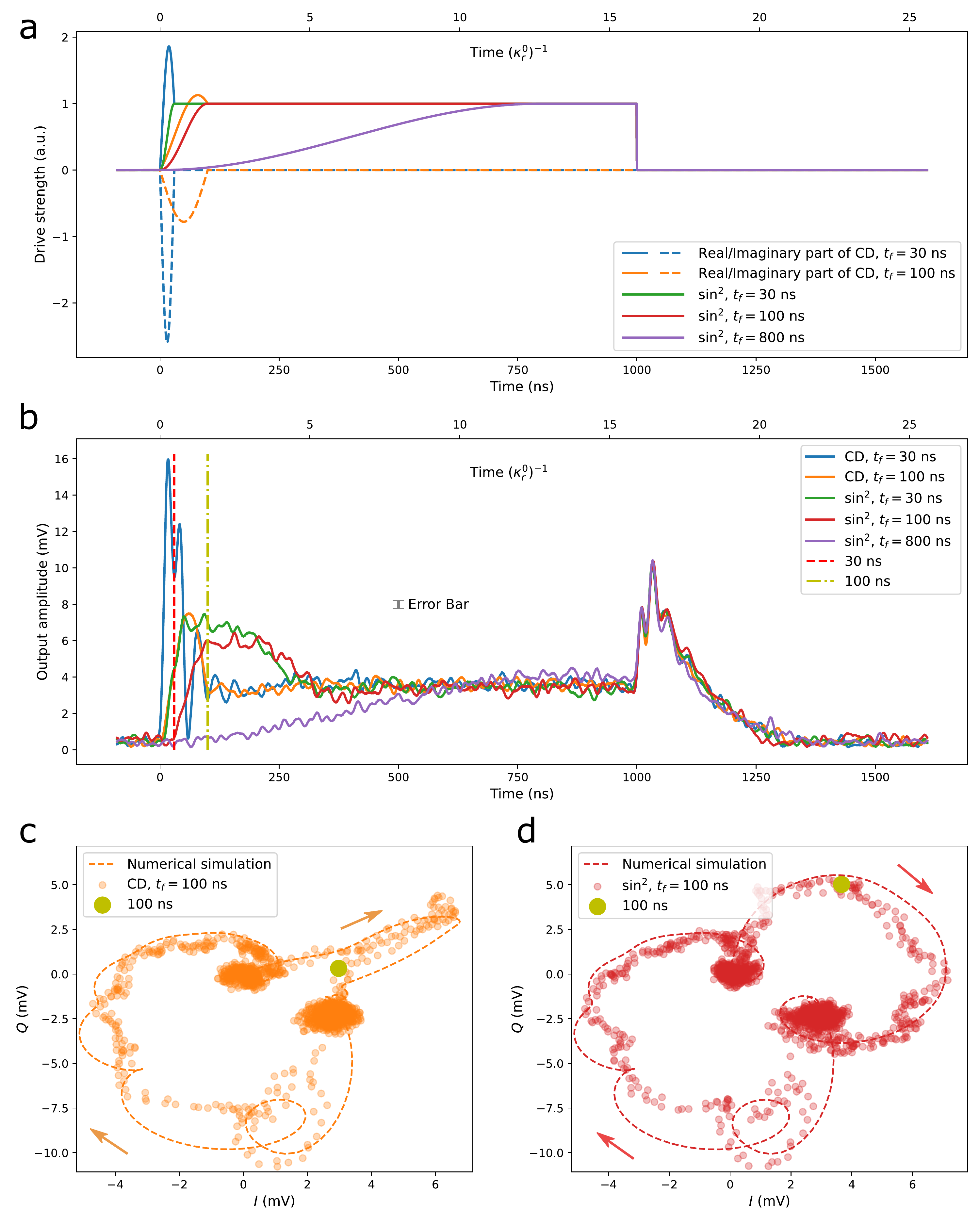}
\end{center}
\caption{\textbf{Open system shortcut to adiabaticity by counterdiabatic (CD) driving and bare drivings.} 
(a) Different pulses used in the experiment. 
The $\sin^2$ ringups with final time $t_f=30\ns$ (green) and $100\ns$ (red) are used, along with their corresponding CD drivings (blue and orange).
The drivings are kept constant at $\epsilon_0$ after $t_f$. 
A slow varying $\sin^2$ driving with $t_f = 800\ns$ (purple) is used to illustrate the adiabatic ringup timescale. 
(b) Output amplitude (in mV) measured by room-temperature FPGA. 
End times of the CD driving protocols are marked for $t_f=30\ns$ (red) and $100\ns$ (yellow). 
The non-equilibrium excitation after $t_f=30 \ns$ is caused by excitation of the Purcell filter mode due to the relatively small resonator-filter detuning. 
The error bar is the standard deviation of the points in the equilibrium states at $990\ns$.
(c,d) IQ trajectories for the $t_f=100\ns$ $\text{sin}^2$ driving (d) and its CD driving (c). 
Arrows alongside the simulation indicate the direction of the time evolution.
The simulated trajectories are calculated based on the input-output formalism detailed in Supplementary Note 1.
To avoid the difficulty of direct tomography, we can infer the bosonic system mean field state from the output signal.
The good fit between simulation and experiment data suggests that the CD driving does indeed realize adiabatic following.
The mismatch mainly comes from the weak nonlinearity of the resonator.  
All experimental points are averaged over $3\times10^4$ experiments. 
A Savitzky–Golay filter with window length 21 and polynomial order 3 is applied to improve the signal-noise ratio.
}
\label{main:cd_result}
\end{figure*}

As shown in Fig.~\ref{main:cd_result}, we apply the reference driving $\epsilon(t)$ with a $\sin^2$-shape ringup, i.e. $\epsilon(t)=\epsilon_0\sin^2(\pi t/2t_f)$ before $t_f$ and $\epsilon(t)=\epsilon_0$ afterwards.
This $\sin^2$-shape reference waveform is chosen to give a smooth, hardware-friendly CD driving pulse.
For both $t_f=30\ns$ and $100\ns$, we compare the performance of the reference driving $\epsilon(t)$ and CD driving $\epsilon_{cd}(t)$. 
To demonstrate a speedup compared with the adiabatic timescale \cite{sarandy2005}, estimated to be on the order of $\kappa_r^{-1}$ (see SN 3), we also apply an adiabatic $\sin^2$ driving.
Until $t_f=800\ns\gg \kappa_r^{-1}$, we observe a relatively good $\sin^2$-shape response, indicating adiabatic evolution.
Our results show an equilibrium time of $100\ns$ for the CD driving and $350\ns \sim 5\kappa_r^{-1}$ for a quench driving.
For $t_f=30\ns$, the large and rapidly varying CD pulse induces out-of-equilibrium excitation of the untargeted mode $a'$.
As a consequence, $a'$ requires an extra relaxation time of $75\ns\sim 5\kappa_f^{-1}$ to return to its equilibrium.
For $t_f=100\ns$, equilibrium is achieved almost immediately after $t_f$, and the non-equilibrium excitation of $a'$ is negligible. 
Since the sampled output signal is a coherent superposition of the input driving field and the leakage of the system (see SN 1 for details), the spiked IQ trajectory during the CD driving in Fig.~\ref{main:cd_result}c does not imply non-equilibrium dynamics.
More data, corresponding to different $t_f$, can be found in SN 9.

The CD drive Eq.~\ref{eq:cd_drive} can only accelerate adiabatic dynamics for a single, qubit-state-dependent hybrid mode at a time and may excite other modes out of their instantaneous eigenstates. This issue can be mitigated with a relatively small driving-detuning ratio for the untargeted modes (Fig.~\ref{main:problem_setup}c).
However, due to the time-energy uncertainty, when the protocol duration is reduced, the correspondingly larger drive amplitudes mean such unwanted excitations cannot be avoided entirely.

\section{Multi-mode optimal control}
To eliminate the excitation of the untargeted mode $a'$ and remove the qubit-state dependence of the driving, we propose the following MMOC protocol.
The MMOC pulse is based on a multi-section ansatz: the protocol duration is divided into $m$ equal-spaced sections, each with a constant complex amplitude, i.e. $\epsilon(t_{j-1}<t<t_j) =\epsilon_{j},\ j\in \{1,2,3,\dots, m\}$. 
During the ringup stage, the control pulse is subject to the boundary conditions $\epsilon(0)=0$ and $ \epsilon(t_f)=\epsilon_0$.
During the reset stage, the boundary conditions are reversed. 
To equilibrate four hybrid modes in time $t_f$, we utilize the underlying Langevin dynamics and obtain four linear equilibrium-transfer equations connecting the pulse vector $\vec{\epsilon}$ and the equilibrium difference $\vec{y}$ with a transfer matrix $G$ as $\vec{y}=G\cdot\vec{\epsilon}$. 
If the complex hybrid detuning  $\Tilde{\Delta} \equiv \Delta - i\kappa/2$ is found for each hybrid mode, $G$ and $\vec{y}$ can be analytically derived. 
These equilibrium-transfer equations can be solved via singular value decomposition of $G$ as long as $m\geq4$.
The resulting $\vec{\epsilon}$ takes the form $\vec{\epsilon}=\vec{\epsilon_e}+\sum_{i=1}^{m-4}x_i\vec{\epsilon_i}$, where the essential vector $\vec{\epsilon_e}$ can be  analytically solved for, and $\vec{\epsilon_i}$ are $m-4$ vectors orthogonal to $\vec{\epsilon_e}$.
See SN 5 for a detailed derivation.
The extra degrees of freedom $x_i$ are chosen numerically to minimize the maximum amplitude component of $\vec{\epsilon}$.
As a result, we can obtain a $5 \dB$ reduction in the maximum amplitude compared with the essential vector. 
See SN 6 for more discussion on the performance and speed limit of this protocol.

\begin{figure*}[tbp]
\begin{center}
\includegraphics[width=0.6\twocolfig]{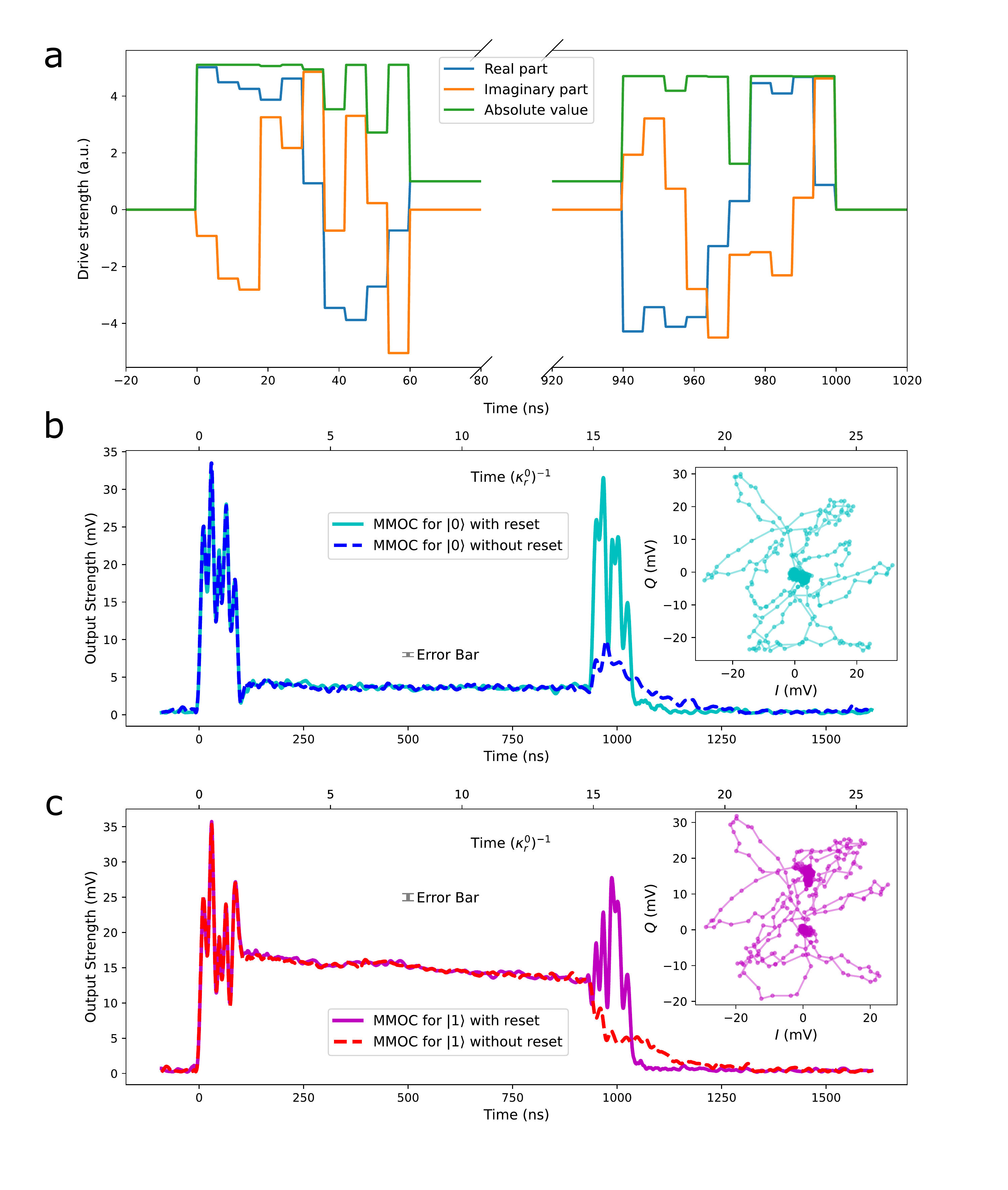}
\end{center}
\caption{\textbf{Multi-mode optimal control (MMOC) by single-port driving.} 
(a) MMOC pulses. 
A multi-section pulse with target equilibrium time $t_f=60\ns$ is applied to shortcut thermal equilibrium of both resonator and filter modes for different qubit states simultaneously. 
A different pulse is used to reset to the vacuum state in the end. 
(b,c) Measured output amplitudes (in mV) for the MMOC pulses with and without reset. 
The steady output signal is achieved for qubit state $\left|0\right>$ (b) and $\left|1\right>$ (c) about $30\ns$ later than the target time, likely due to the high driving amplitude and low-Q energy-storing components in the feedline, such as the impedance-matched Josephson parametric amplifier.
This explanation is reinforced by the observation of a similar $30\ns$ tail with a far-detuned and high-amplitude driving which, in principle, will not excite the multi-mode cQED system (SN 9).
In (c), the steady output decays over time due to the $T_1$ decay of the qubit.
The error bar is the standard deviation of the points in the equilibrium states at $930\ns$.
The good final reset performance for the mixed qubit state demonstrates that MMOC works for both qubit states simultaneously. 
The corresponding IQ trajectories are plotted inset, from which we can infer the system undergoes highly non-equilibrium dynamics during the MMOC pulse.
All data are processed in the same way as in Fig.~\ref{main:cd_result}.
}
\label{main:oc_result}
\end{figure*}

Fig.~\ref{main:oc_result} shows our results for the fast equilibration of all four lossy hybrid normal modes of the coupled oscillator system. 
Choosing $m=10$, we first apply a $t_f = 60\ns$ ringup pulse for fast system equilibration, and then transfer the system to the vacuum state at the end with another $60\ns$ reset pulse.
According to the time-traced IQ trajectories (Fig.~\ref{main:oc_result}b, c and insets), different modes undergo different non-equilibrium dynamics and end up in their respective equilibrium states after the MMOC pulse.
Our results show that fast unconditional multi-mode ringup and depletion are almost achieved at the desired time and reduce the duration required to achieve equilibrium compared with natural relaxation.

\section{Discussion}
We have experimentally extended STA to an open quantum system.
Our CD method accelerates adiabatic dynamics of the DFS for a single driven-dissipative bosonic mode to occur within $100 \ns$, compared with the $800 \ns$ of the regular slow varying scheme.
Furthermore, by utilizing possible non-adiabatic dynamics, our MMOC protocol -- based on methods from optimal control theory -- can achieve unconditional adiabatic dynamics for multiple lossy bosonic modes simultaneously in a similar duration. 
It is worth pointing out that the ringdown pulse cannot be obtained by simply reversing the ringup pulse as can be done in the closed system.
Time-reversal symmetry is broken under open-system dynamics.

\begin{figure*}[tbp]
\begin{center}
\includegraphics[width=0.5\twocolfig]{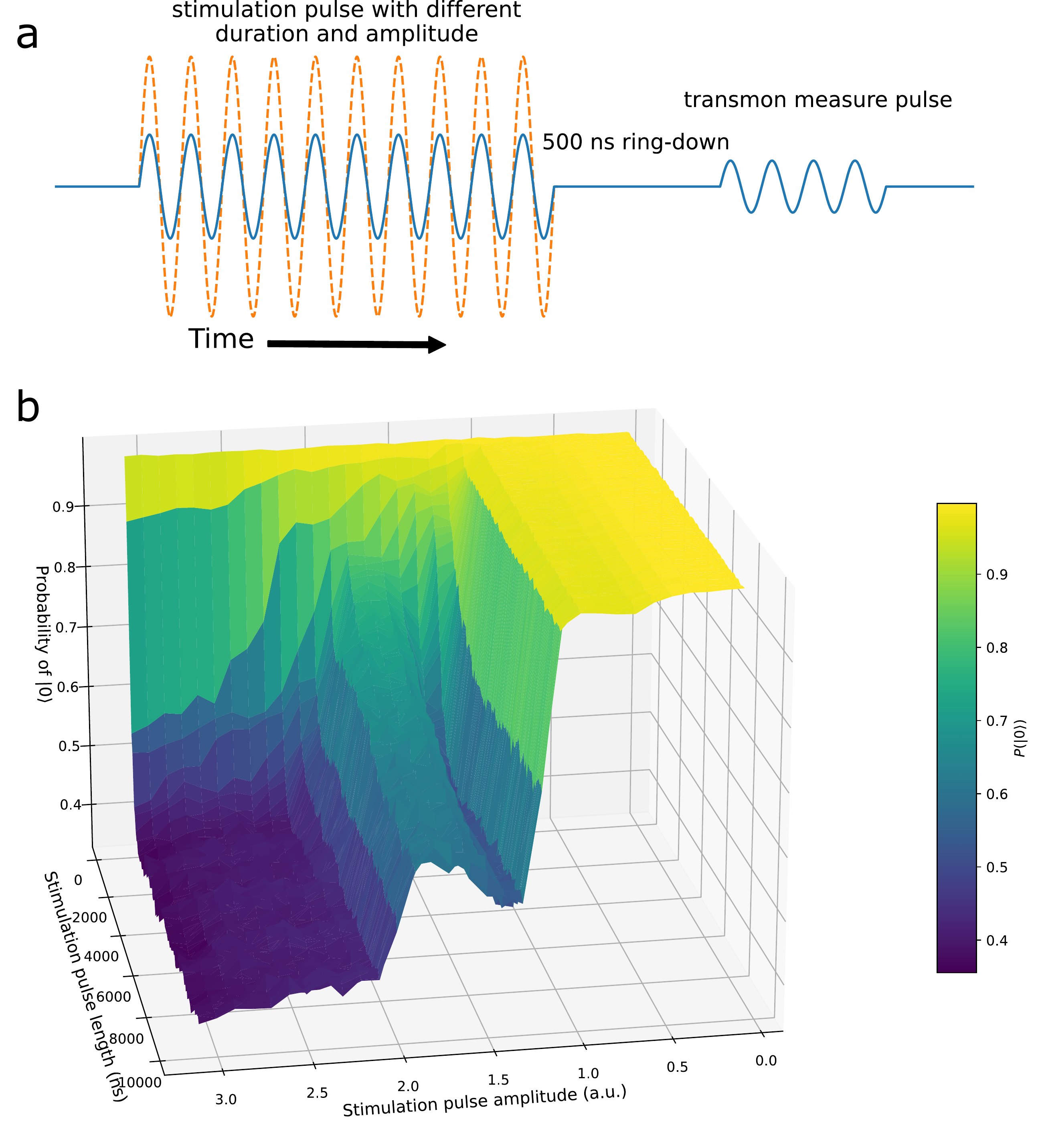}
\end{center}
\caption{\textbf{Qubit excitation by feedline driving.} 
(a) The pulse sequence used to check the impact of the microwave driving from the feedline on the transmon qubit.
We apply a stimulating microwave pulse with fixed frequency $\omega_d$ and different amplitudes and durations.
We wait for $500\ns$ to clear the resonator-filter population before performing the qubit dispersive measurement.
(b) Population of $\left|0\right>$ as a function of the amplitude and duration of the stimulating pulse.
Once a critical pulse amplitude is exceeded, the qubit is excited. 
Our results are qualitatively similar to previous findings~\cite{sank2016measurement}.
To avoid unwanted excitation, we optimize the driving pulse amplitude and the speed of the protocol. 
For other linear lossy bosonic mode systems without a dispersively coupled qubit (e.g. optomechanics), STA will not be limited in this way. 
}
\label{main:me_result}
\end{figure*}

One issue preventing our protocols from further acceleration is the Kerr nonlinearity correction term $\frac{1}{2}K(b'^{\dagger})^2 b'^2$ to the resonator mode dispersive Hamiltonian~\cite{blais2020circuit}.
While our methods are designed to work in the linear regime, increasing the protocol speed requires fast-growing driving power, and accounting for nonlinearity becomes essential.
However, in the weakly nonlinear regime, i.e., where driving power is well below the first-order dissipative phase transition point~\cite{brookes2021critical}, we empirically, in both simulation and experiment, find the effect of nonlinearity can be largely mitigated by including a mean resonator frequency shift $\sim K n$ \cite{leghtas2015confining} in our protocols, where $n$ is the equilibrium photon number in the resonator mode.
If the drive exceeds this dissipative phase transition point, not only does the microwave output suddenly increase, we also observe the transmon qubit becoming excited in a qualitatively similar way to a previously reported result~\cite{sank2016measurement}.
This result is shown in Fig.~\ref{main:me_result}, where we stimulate the resonator mode with different amplitudes and durations and measure the remaining ground state population.
Our results support the theory \cite{mavrogordatos2017simultaneous} that the resonator phase transition and the qubit excitation coincide.

The methods we have presented here are a first step to accelerating open system dynamics in a broader range of settings.
The open system CD suppresses diabatic transitions out of the DFS and opens the door for STA to more applications, such as quantum thermodynamics, reservoir engineering~\cite{albert2016geometry}, and holonomic quantum computation based on bosonic codes~\cite{albert2016holonomic}.
The open-loop MMOC protocol, being independent of qubit-state, is suitable for fast resonator ringup(ringdown)~\cite{mcclure2016rapid,bultink2016active} and can be used to reduce measurement cycle times in quantum error correction protocols.
Both methods can also potentially be applied to other dissipative bosonic systems, such as optical resonant cavities and optomechanics.
As recent theoretical developments have shown, open-system STA also has interdisciplinary applications such as accelerating biophysics~\cite{chen2021dynamics}, bioengineering~\cite{iram2021controlling},  and chemical reaction dynamics~\cite{koch2016controlling}.
\section{Methods}
\noindent\textbf{Experimental Setup and Calibration.}
Our devices consists of a tunable transmon qubit coupled to a $\lambda/4$ microwave resonator with coupling $g/2\pi \approx 80 \MHz$. 
The resonator is coupled to an individual $\lambda/4$ Purcell filter with coupling $J/2\pi \approx 10\MHz$. 
The hybrid mode frequencies are $\omega^0_r/2\pi = 6.4427\GHz,\ \omega^1_r/2\pi = 6.4378\GHz$ and $\omega^0_f/2\pi = 6.4634\GHz,\ \omega^1_f/2\pi = 6.4673\GHz$ and linewidths $\kappa^0_r = 1/62.88\ns,\ \kappa^1_r = 1/77.93\ns$ and $\kappa^0_f = 1/17.86\ns,\ \kappa^1_f = 1/15.64\ns$. 
The frequencies $\omega_r,\ \omega_f$ can be estimated by fitting the transmission $S_{21}$ shown in Fig.~\ref{main:problem_setup}c with the input-output theory of SN~1, or by fitting the measured time-traced IQ plots with numerical trajectories obtained from quantum Langevin dynamics. 
The hybrid linewidths $\kappa_r,\ \kappa_f$ are obtained by measuring the decay time constant of the quenched ringdown. 
All spectrum parameters are characterized to a precision of $< 2\pi\times0.1\MHz$.

Driving pulses $\epsilon(t)$ are generated by an arbitrary waveform generator with a sampling rate of $2\unit{GS}/s$ and up-converted to the driving frequency by IQ mixing with the LO microwave signal. 
The readout signal is first amplified with an impedance-matched Josephson parametric amplifier, then by high electron mobility transistors, and finally by room temperature amplifiers. 
It is further homodyne demodulated, digitized by an analogue-to-digital converter at $1\unit{GS}/s$ and analyzed by a room-temperature DAQ FPGA.
Each experimental point in the figures is an average of $3\times 10^4$ experiments.

\noindent\textbf{Hybrid-Mode Dynamics.}
Here we show how the interacting dissipative modes are decoupled so that CD driving is realized for a single lossy hybrid mode. 
The derivation is based on a simplified version of the circuit model shown in Fig.~\ref{main:problem_setup}b, where the input gate capacitor is ignored, and the qubit's effect is implicitly accounted for through a state-dependent shift on the filter and resonator modes.
The exact input-output relations based on the entire circuit are derived in SN 1, which are equivalent to the results of this simplified model through a parameter renormalization.

In the dispersive regime of cQED~\cite{blais2004cavity}, after applying the rotating wave approximation, the bare system Hamiltonian in the driving frame (at frequency $\omega_d$) reads
\begin{linenomath*}
\begin{equation}
    H^{0,1} = \Delta_a a^\dagger a + \Delta_b^{0,1} b^\dagger b + J(a^\dagger b + b^\dagger a),
    \label{eq:sys_H}
\end{equation}
\end{linenomath*}
where $a,\ b$ are bare filter and resonator modes, the superscripts denote the qubit state, $\Delta_{a(b)}\equiv\omega_{a(b)}-\omega_d$ is the driving detuning of mode $a$($b$), $J$ is coupling strength between the modes, and the qubit-induced dispersive shift is $2\chi = \Delta_b^1 - \Delta_b^0$.
Here we have set $\hbar=1$.
Since our protocol time is much shorter than the qubit lifetime, decoherence is dominated by $\kappa_a$ through the cold bath at an effective temperature of $75\mK$. 
In the Heisenberg picture, the system dynamics according to the input-output theory is \cite{gardiner1985}
\begin{linenomath*}
\begin{align}
    \dot{a} &= -i\Delta_a a - \frac{\kappa_a}{2} a - i J b - \sqrt{\kappa_a}a_i\label{eq:langevin_a}\\
    \dot{b} &= -i\Delta^{0,1}_b b - i J a
    \label{eq:langevin_b},
\end{align}
\end{linenomath*}
where $a_i$ is the time-dependent input field. 
A linear transformation of $a$ and $b$ decouples these equations, resulting in the hybrid modes $a'^{0,1},\ b'^{0,1}$ with frequencies $\Delta_f^{0,1},\ \Delta_r^{0,1}$ and linewidths $\kappa_f^{0,1},\ \kappa_r^{0,1}$. 
The hybrid modes are defined so that $[a'^i,a'^{i\dagger}]=[b'^i,b'^{i\dagger}]=1$ for $i=0,1$, whereas $[a'^i,b'^i]\neq 0$ due to environmental effects.
Omitting the superscripts for convenience, we have
\begin{linenomath*}
\begin{align}
    \dot{a}' &= -i\Delta_f a' - \frac{\kappa_f}{2}a' - c_f\sqrt{\kappa_a}a_i
    \label{eq:langevin_f}\\
    \dot{b}' &= -i\Delta_r b' - \frac{\kappa_r}{2}b' - c_r\sqrt{\kappa_a}a_i,
    \label{eq:langevin_r}
\end{align}
\end{linenomath*}
where $c_f,\ c_r$ are coefficients from the linear transformation.
Eq.~\ref{eq:langevin_r} is equivalent to the master equation in the Schrodinger picture~\cite{breuer2002theory}:
\begin{equation}
    \dot{\rho}_{b'} = -i[H_r(t), \rho_{b'}] + \kappa_r \mathcal{D}(b')\rho_{b'},
    \label{eq:adj_me_r}
\end{equation}
where $H_r(t)=\Delta_r b'^\dagger b' -(i c_r\sqrt{\kappa_a}a_i(t) b'^\dagger + h.c.)$.
For coherent driving fields, we can approximate the Hamiltonian as $H_r(t)=\Delta_r b'^\dagger b' - (i \epsilon_r(t) b'^\dagger + h.c.)$, where the effective drive is $\epsilon_r(t)=c_r\sqrt{\kappa_a}\langle a_i(t)\rangle$.
Based on the master equation \ref{eq:adj_me_r}, we can derive (see SN 3) the CD driving Eq.~\eqref{eq:cd_drive} for a single hybrid mode. 

\noindent\textbf{Data availability.}\\
Raw data for the results reported in the text are available
from the authors upon request.
\bibliography{MRF}

\begin{thebibliography}{10}
\expandafter\ifx\csname url\endcsname\relax
  \def\url#1{\texttt{#1}}\fi
\expandafter\ifx\csname urlprefix\endcsname\relax\def\urlprefix{URL }\fi
\providecommand{\bibinfo}[2]{#2}
\providecommand{\eprint}[2][]{\url{#2}}

\bibitem{farhi2001quantum}
\bibinfo{author}{Farhi, E.} \emph{et~al.}
\newblock \bibinfo{title}{A quantum adiabatic evolution algorithm applied to
  random instances of an np-complete problem}.
\newblock \emph{\bibinfo{journal}{Science}} \textbf{\bibinfo{volume}{292}},
  \bibinfo{pages}{472--475} (\bibinfo{year}{2001}).

\bibitem{aharonov2008adiabatic}
\bibinfo{author}{Aharonov, D.} \emph{et~al.}
\newblock \bibinfo{title}{Adiabatic quantum computation is equivalent to
  standard quantum computation}.
\newblock \emph{\bibinfo{journal}{SIAM review}} \textbf{\bibinfo{volume}{50}},
  \bibinfo{pages}{755--787} (\bibinfo{year}{2008}).

\bibitem{zanardi1999holonomic}
\bibinfo{author}{Zanardi, P.} \& \bibinfo{author}{Rasetti, M.}
\newblock \bibinfo{title}{Holonomic quantum computation}.
\newblock \emph{\bibinfo{journal}{Physics Letters A}}
  \textbf{\bibinfo{volume}{264}}, \bibinfo{pages}{94--99}
  (\bibinfo{year}{1999}).

\bibitem{ekert2000geometric}
\bibinfo{author}{Ekert, A.} \emph{et~al.}
\newblock \bibinfo{title}{Geometric quantum computation}.
\newblock \emph{\bibinfo{journal}{Journal of Modern Optics}}
  \textbf{\bibinfo{volume}{47}}, \bibinfo{pages}{2501--2513}
  (\bibinfo{year}{2000}).

\bibitem{childs2001robustness}
\bibinfo{author}{Childs, A.~M.}, \bibinfo{author}{Farhi, E.} \&
  \bibinfo{author}{Preskill, J.}
\newblock \bibinfo{title}{Robustness of adiabatic quantum computation}.
\newblock \emph{\bibinfo{journal}{Physical Review A}}
  \textbf{\bibinfo{volume}{65}}, \bibinfo{pages}{012322}
  (\bibinfo{year}{2001}).

\bibitem{chen2010fast}
\bibinfo{author}{Chen, X.} \emph{et~al.}
\newblock \bibinfo{title}{Fast optimal frictionless atom cooling in harmonic
  traps: Shortcut to adiabaticity}.
\newblock \emph{\bibinfo{journal}{Physical review letters}}
  \textbf{\bibinfo{volume}{104}}, \bibinfo{pages}{063002}
  (\bibinfo{year}{2010}).

\bibitem{guery2014transport}
\bibinfo{author}{Gu{\'e}ry-Odelin, D.} \& \bibinfo{author}{Muga, J.}
\newblock \bibinfo{title}{Transport in a harmonic trap: Shortcuts to
  adiabaticity and robust protocols}.
\newblock \emph{\bibinfo{journal}{Physical Review A}}
  \textbf{\bibinfo{volume}{90}}, \bibinfo{pages}{063425}
  (\bibinfo{year}{2014}).

\bibitem{an2016shortcuts}
\bibinfo{author}{An, S.}, \bibinfo{author}{Lv, D.}, \bibinfo{author}{Del~Campo,
  A.} \& \bibinfo{author}{Kim, K.}
\newblock \bibinfo{title}{Shortcuts to adiabaticity by counterdiabatic driving
  for trapped-ion displacement in phase space}.
\newblock \emph{\bibinfo{journal}{Nature communications}}
  \textbf{\bibinfo{volume}{7}}, \bibinfo{pages}{1--5} (\bibinfo{year}{2016}).

\bibitem{bason2012high}
\bibinfo{author}{Bason, M.~G.} \emph{et~al.}
\newblock \bibinfo{title}{High-fidelity quantum driving}.
\newblock \emph{\bibinfo{journal}{Nature Physics}}
  \textbf{\bibinfo{volume}{8}}, \bibinfo{pages}{147--152}
  (\bibinfo{year}{2012}).

\bibitem{zhang2013experimental}
\bibinfo{author}{Zhang, J.} \emph{et~al.}
\newblock \bibinfo{title}{Experimental implementation of assisted quantum
  adiabatic passage in a single spin}.
\newblock \emph{\bibinfo{journal}{Physical review letters}}
  \textbf{\bibinfo{volume}{110}}, \bibinfo{pages}{240501}
  (\bibinfo{year}{2013}).

\bibitem{du2016experimental}
\bibinfo{author}{Du, Y.-X.} \emph{et~al.}
\newblock \bibinfo{title}{Experimental realization of stimulated raman
  shortcut-to-adiabatic passage with cold atoms}.
\newblock \emph{\bibinfo{journal}{Nature communications}}
  \textbf{\bibinfo{volume}{7}}, \bibinfo{pages}{1--7} (\bibinfo{year}{2016}).

\bibitem{zhou2017accelerated}
\bibinfo{author}{Zhou, B.~B.} \emph{et~al.}
\newblock \bibinfo{title}{Accelerated quantum control using superadiabatic
  dynamics in a solid-state lambda system}.
\newblock \emph{\bibinfo{journal}{Nature Physics}}
  \textbf{\bibinfo{volume}{13}}, \bibinfo{pages}{330--334}
  (\bibinfo{year}{2017}).

\bibitem{martinis2014fast}
\bibinfo{author}{Martinis, J.~M.} \& \bibinfo{author}{Geller, M.~R.}
\newblock \bibinfo{title}{Fast adiabatic qubit gates using only $\sigma$ z
  control}.
\newblock \emph{\bibinfo{journal}{Physical Review A}}
  \textbf{\bibinfo{volume}{90}}, \bibinfo{pages}{022307}
  (\bibinfo{year}{2014}).

\bibitem{theis2018counteracting}
\bibinfo{author}{Theis, L.}, \bibinfo{author}{Motzoi, F.},
  \bibinfo{author}{Machnes, S.} \& \bibinfo{author}{Wilhelm, F.}
\newblock \bibinfo{title}{Counteracting systems of diabaticities using drag
  controls: The status after 10 years (a)}.
\newblock \emph{\bibinfo{journal}{EPL (Europhysics Letters)}}
  \textbf{\bibinfo{volume}{123}}, \bibinfo{pages}{60001}
  (\bibinfo{year}{2018}).

\bibitem{an2015experimental}
\bibinfo{author}{An, S.} \emph{et~al.}
\newblock \bibinfo{title}{Experimental test of the quantum jarzynski equality
  with a trapped-ion system}.
\newblock \emph{\bibinfo{journal}{Nature Physics}}
  \textbf{\bibinfo{volume}{11}}, \bibinfo{pages}{193--199}
  (\bibinfo{year}{2015}).

\bibitem{guery2019shortcuts}
\bibinfo{author}{Gu{\'e}ry-Odelin, D.} \emph{et~al.}
\newblock \bibinfo{title}{Shortcuts to adiabaticity: Concepts, methods, and
  applications}.
\newblock \emph{\bibinfo{journal}{Reviews of Modern Physics}}
  \textbf{\bibinfo{volume}{91}}, \bibinfo{pages}{045001}
  (\bibinfo{year}{2019}).

\bibitem{demirplak2003adiabatic}
\bibinfo{author}{Demirplak, M.} \& \bibinfo{author}{Rice, S.~A.}
\newblock \bibinfo{title}{Adiabatic population transfer with control fields}.
\newblock \emph{\bibinfo{journal}{The Journal of Physical Chemistry A}}
  \textbf{\bibinfo{volume}{107}}, \bibinfo{pages}{9937--9945}
  (\bibinfo{year}{2003}).

\bibitem{berry2009transitionless}
\bibinfo{author}{Berry, M.~V.}
\newblock \bibinfo{title}{Transitionless quantum driving}.
\newblock \emph{\bibinfo{journal}{Journal of Physics A: Mathematical and
  Theoretical}} \textbf{\bibinfo{volume}{42}}, \bibinfo{pages}{365303}
  (\bibinfo{year}{2009}).

\bibitem{song2016shortcuts}
\bibinfo{author}{Song, X.-K.}, \bibinfo{author}{Zhang, H.},
  \bibinfo{author}{Ai, Q.}, \bibinfo{author}{Qiu, J.} \& \bibinfo{author}{Deng,
  F.-G.}
\newblock \bibinfo{title}{Shortcuts to adiabatic holonomic quantum computation
  in decoherence-free subspace with transitionless quantum driving algorithm}.
\newblock \emph{\bibinfo{journal}{New Journal of Physics}}
  \textbf{\bibinfo{volume}{18}}, \bibinfo{pages}{023001}
  (\bibinfo{year}{2016}).

\bibitem{beau2016scaling}
\bibinfo{author}{Beau, M.}, \bibinfo{author}{Jaramillo, J.} \&
  \bibinfo{author}{Del~Campo, A.}
\newblock \bibinfo{title}{Scaling-up quantum heat engines efficiently via
  shortcuts to adiabaticity}.
\newblock \emph{\bibinfo{journal}{Entropy}} \textbf{\bibinfo{volume}{18}},
  \bibinfo{pages}{168} (\bibinfo{year}{2016}).

\bibitem{martinez2016engineered}
\bibinfo{author}{Mart{\'\i}nez, I.~A.}, \bibinfo{author}{Petrosyan, A.},
  \bibinfo{author}{Gu{\'e}ry-Odelin, D.}, \bibinfo{author}{Trizac, E.} \&
  \bibinfo{author}{Ciliberto, S.}
\newblock \bibinfo{title}{Engineered swift equilibration of a brownian
  particle}.
\newblock \emph{\bibinfo{journal}{Nature physics}}
  \textbf{\bibinfo{volume}{12}}, \bibinfo{pages}{843--846}
  (\bibinfo{year}{2016}).

\bibitem{iram2021controlling}
\bibinfo{author}{Iram, S.} \emph{et~al.}
\newblock \bibinfo{title}{Controlling the speed and trajectory of evolution
  with counterdiabatic driving}.
\newblock \emph{\bibinfo{journal}{Nature Physics}}
  \textbf{\bibinfo{volume}{17}}, \bibinfo{pages}{135--142}
  (\bibinfo{year}{2021}).

\bibitem{ibanez2011shortcuts}
\bibinfo{author}{Ib{\'a}{\~n}ez, S.}, \bibinfo{author}{Mart{\'\i}nez-Garaot,
  S.}, \bibinfo{author}{Chen, X.}, \bibinfo{author}{Torrontegui, E.} \&
  \bibinfo{author}{Muga, J.~G.}
\newblock \bibinfo{title}{Shortcuts to adiabaticity for non-hermitian systems}.
\newblock \emph{\bibinfo{journal}{Physical Review A}}
  \textbf{\bibinfo{volume}{84}}, \bibinfo{pages}{023415}
  (\bibinfo{year}{2011}).

\bibitem{chen2016method}
\bibinfo{author}{Chen, Y.-H.}, \bibinfo{author}{Xia, Y.}, \bibinfo{author}{Wu,
  Q.-C.}, \bibinfo{author}{Huang, B.-H.} \& \bibinfo{author}{Song, J.}
\newblock \bibinfo{title}{Method for constructing shortcuts to adiabaticity by
  a substitute of counterdiabatic driving terms}.
\newblock \emph{\bibinfo{journal}{Physical Review A}}
  \textbf{\bibinfo{volume}{93}}, \bibinfo{pages}{052109}
  (\bibinfo{year}{2016}).

\bibitem{vacanti2014}
\bibinfo{author}{Vacanti, G.} \emph{et~al.}
\newblock \bibinfo{title}{Transitionless quantum driving in open quantum
  systems}.
\newblock \emph{\bibinfo{journal}{New Journal of Physics}}
  \textbf{\bibinfo{volume}{16}}, \bibinfo{pages}{053017}
  (\bibinfo{year}{2014}).

\bibitem{villazon2019swift}
\bibinfo{author}{Villazon, T.}, \bibinfo{author}{Polkovnikov, A.} \&
  \bibinfo{author}{Chandran, A.}
\newblock \bibinfo{title}{Swift heat transfer by fast-forward driving in open
  quantum systems}.
\newblock \emph{\bibinfo{journal}{Physical Review A}}
  \textbf{\bibinfo{volume}{100}}, \bibinfo{pages}{012126}
  (\bibinfo{year}{2019}).

\bibitem{sarandy2005adiabatic}
\bibinfo{author}{Sarandy, M.} \& \bibinfo{author}{Lidar, D.}
\newblock \bibinfo{title}{Adiabatic approximation in open quantum systems}.
\newblock \emph{\bibinfo{journal}{Physical Review A}}
  \textbf{\bibinfo{volume}{71}}, \bibinfo{pages}{012331}
  (\bibinfo{year}{2005}).

\bibitem{lidar1998decoherence}
\bibinfo{author}{Lidar, D.~A.}, \bibinfo{author}{Chuang, I.~L.} \&
  \bibinfo{author}{Whaley, K.~B.}
\newblock \bibinfo{title}{Decoherence-free subspaces for quantum computation}.
\newblock \emph{\bibinfo{journal}{Physical Review Letters}}
  \textbf{\bibinfo{volume}{81}}, \bibinfo{pages}{2594} (\bibinfo{year}{1998}).

\bibitem{wu2017adiabatic}
\bibinfo{author}{Wu, S.}, \bibinfo{author}{Huang, X.}, \bibinfo{author}{Li,
  H.}, \bibinfo{author}{Yi, X.} \emph{et~al.}
\newblock \bibinfo{title}{Adiabatic evolution of decoherence-free subspaces and
  its shortcuts}.
\newblock \emph{\bibinfo{journal}{Physical Review A}}
  \textbf{\bibinfo{volume}{96}}, \bibinfo{pages}{042104}
  (\bibinfo{year}{2017}).

\bibitem{sank2016measurement}
\bibinfo{author}{Sank, D.} \emph{et~al.}
\newblock \bibinfo{title}{Measurement-induced state transitions in a
  superconducting qubit: Beyond the rotating wave approximation}.
\newblock \emph{\bibinfo{journal}{Physical review letters}}
  \textbf{\bibinfo{volume}{117}}, \bibinfo{pages}{190503}
  (\bibinfo{year}{2016}).

\bibitem{zhou2021rapid}
\bibinfo{author}{Zhou, Y.} \emph{et~al.}
\newblock \bibinfo{title}{Rapid and unconditional parametric reset protocol for
  tunable superconducting qubits}.
\newblock \emph{\bibinfo{journal}{arXiv preprint arXiv:2103.11315}}
  (\bibinfo{year}{2021}).

\bibitem{koch2007transmon}
\bibinfo{author}{Koch, J.} \emph{et~al.}
\newblock \bibinfo{title}{Charge-insensitive qubit design derived from the
  cooper pair box}.
\newblock \emph{\bibinfo{journal}{Physical Review A}}
  \textbf{\bibinfo{volume}{76}}, \bibinfo{pages}{042319}
  (\bibinfo{year}{2007}).

\bibitem{taddei2013quantum}
\bibinfo{author}{Taddei, M.~M.}, \bibinfo{author}{Escher, B.~M.},
  \bibinfo{author}{Davidovich, L.} \& \bibinfo{author}{de~Matos~Filho, R.~L.}
\newblock \bibinfo{title}{Quantum speed limit for physical processes}.
\newblock \emph{\bibinfo{journal}{Physical review letters}}
  \textbf{\bibinfo{volume}{110}}, \bibinfo{pages}{050402}
  (\bibinfo{year}{2013}).

\bibitem{sarandy2005}
\bibinfo{author}{Sarandy, M.} \& \bibinfo{author}{Lidar, D.}
\newblock \bibinfo{title}{Adiabatic approximation in open quantum systems}.
\newblock \emph{\bibinfo{journal}{Physical Review A}}
  \textbf{\bibinfo{volume}{71}}, \bibinfo{pages}{012331}
  (\bibinfo{year}{2005}).

\bibitem{blais2020circuit}
\bibinfo{author}{Blais, A.}, \bibinfo{author}{Grimsmo, A.~L.},
  \bibinfo{author}{Girvin, S.~M.} \& \bibinfo{author}{Wallraff, A.}
\newblock \bibinfo{title}{Circuit quantum electrodynamics}.
\newblock \emph{\bibinfo{journal}{Rev. Mod. Phys.}}
  \textbf{\bibinfo{volume}{93}}, \bibinfo{pages}{025005}
  (\bibinfo{year}{2021}).

\bibitem{brookes2021critical}
\bibinfo{author}{Brookes, P.} \emph{et~al.}
\newblock \bibinfo{title}{Critical slowing down in circuit quantum
  electrodynamics}.
\newblock \emph{\bibinfo{journal}{Science Advances}}
  \textbf{\bibinfo{volume}{7}}, \bibinfo{pages}{eabe9492}
  (\bibinfo{year}{2021}).

\bibitem{leghtas2015confining}
\bibinfo{author}{Leghtas, Z.} \emph{et~al.}
\newblock \bibinfo{title}{Confining the state of light to a quantum manifold by
  engineered two-photon loss}.
\newblock \emph{\bibinfo{journal}{Science}} \textbf{\bibinfo{volume}{347}},
  \bibinfo{pages}{853--857} (\bibinfo{year}{2015}).

\bibitem{mavrogordatos2017simultaneous}
\bibinfo{author}{Mavrogordatos, T.~K.} \emph{et~al.}
\newblock \bibinfo{title}{Simultaneous bistability of a qubit and resonator in
  circuit quantum electrodynamics}.
\newblock \emph{\bibinfo{journal}{Physical review letters}}
  \textbf{\bibinfo{volume}{118}}, \bibinfo{pages}{040402}
  (\bibinfo{year}{2017}).

\bibitem{albert2016geometry}
\bibinfo{author}{Albert, V.~V.}, \bibinfo{author}{Bradlyn, B.},
  \bibinfo{author}{Fraas, M.} \& \bibinfo{author}{Jiang, L.}
\newblock \bibinfo{title}{Geometry and response of lindbladians}.
\newblock \emph{\bibinfo{journal}{Physical Review X}}
  \textbf{\bibinfo{volume}{6}}, \bibinfo{pages}{041031} (\bibinfo{year}{2016}).

\bibitem{albert2016holonomic}
\bibinfo{author}{Albert, V.~V.} \emph{et~al.}
\newblock \bibinfo{title}{Holonomic quantum control with continuous variable
  systems}.
\newblock \emph{\bibinfo{journal}{Physical review letters}}
  \textbf{\bibinfo{volume}{116}}, \bibinfo{pages}{140502}
  (\bibinfo{year}{2016}).

\bibitem{mcclure2016rapid}
\bibinfo{author}{McClure, D.~T.} \emph{et~al.}
\newblock \bibinfo{title}{Rapid driven reset of a qubit readout resonator}.
\newblock \emph{\bibinfo{journal}{Physical Review Applied}}
  \textbf{\bibinfo{volume}{5}}, \bibinfo{pages}{011001} (\bibinfo{year}{2016}).

\bibitem{bultink2016active}
\bibinfo{author}{Bultink, C.~C.} \emph{et~al.}
\newblock \bibinfo{title}{Active resonator reset in the nonlinear dispersive
  regime of circuit qed}.
\newblock \emph{\bibinfo{journal}{Physical Review Applied}}
  \textbf{\bibinfo{volume}{6}}, \bibinfo{pages}{034008} (\bibinfo{year}{2016}).

\bibitem{chen2021dynamics}
\bibinfo{author}{Chen, K.} \emph{et~al.}
\newblock \bibinfo{title}{Dynamics of driven polymer transport through a
  nanopore}.
\newblock \emph{\bibinfo{journal}{Nature Physics}} \bibinfo{pages}{1--7}
  (\bibinfo{year}{2021}).

\bibitem{koch2016controlling}
\bibinfo{author}{Koch, C.~P.}
\newblock \bibinfo{title}{Controlling open quantum systems: tools,
  achievements, and limitations}.
\newblock \emph{\bibinfo{journal}{Journal of Physics: Condensed Matter}}
  \textbf{\bibinfo{volume}{28}}, \bibinfo{pages}{213001}
  (\bibinfo{year}{2016}).

\bibitem{blais2004cavity}
\bibinfo{author}{Blais, A.}, \bibinfo{author}{Huang, R.-S.},
  \bibinfo{author}{Wallraff, A.}, \bibinfo{author}{Girvin, S.~M.} \&
  \bibinfo{author}{Schoelkopf, R.~J.}
\newblock \bibinfo{title}{Cavity quantum electrodynamics for superconducting
  electrical circuits: An architecture for quantum computation}.
\newblock \emph{\bibinfo{journal}{Physical Review A}}
  \textbf{\bibinfo{volume}{69}}, \bibinfo{pages}{062320}
  (\bibinfo{year}{2004}).

\bibitem{gardiner1985}
\bibinfo{author}{Gardiner, C.~W.} \& \bibinfo{author}{Collett, M.~J.}
\newblock \bibinfo{title}{Input and output in damped quantum systems: Quantum
  stochastic differential equations and the master equation}.
\newblock \emph{\bibinfo{journal}{Physical Review A}}
  \textbf{\bibinfo{volume}{31}}, \bibinfo{pages}{3761} (\bibinfo{year}{1985}).

\bibitem{breuer2002theory}
\bibinfo{author}{Breuer, H.-P.}, \bibinfo{author}{Petruccione, F.}
  \emph{et~al.}
\newblock \emph{\bibinfo{title}{The theory of open quantum systems}}
  (\bibinfo{publisher}{Oxford University Press on Demand},
  \bibinfo{year}{2002}).

\bibitem{heinsoo2018}
\bibinfo{author}{Heinsoo, J.} \emph{et~al.}
\newblock \bibinfo{title}{Rapid high-fidelity multiplexed readout of
  superconducting qubits}.
\newblock \emph{\bibinfo{journal}{Physical Review Applied}}
  \textbf{\bibinfo{volume}{10}}, \bibinfo{pages}{034040}
  (\bibinfo{year}{2018}).

\bibitem{walls2007}
\bibinfo{author}{Walls, D.~F.} \& \bibinfo{author}{Milburn, G.~J.}
\newblock \emph{\bibinfo{title}{Quantum optics}} (\bibinfo{publisher}{Springer
  Science \& Business Media}, \bibinfo{year}{2007}).

\bibitem{bures1969}
\bibinfo{author}{Bures, D.}
\newblock \bibinfo{title}{An extension of kakutani's theorem on infinite
  product measures to the tensor product of semifinite w*-algebras}.
\newblock \emph{\bibinfo{journal}{Transactions of the American Mathematical
  Society}} \textbf{\bibinfo{volume}{135}}, \bibinfo{pages}{199--212}
  (\bibinfo{year}{1969}).

\bibitem{braunstein1994statistical}
\bibinfo{author}{Braunstein, S.~L.} \& \bibinfo{author}{Caves, C.~M.}
\newblock \bibinfo{title}{Statistical distance and the geometry of quantum
  states}.
\newblock \emph{\bibinfo{journal}{Physical Review Letters}}
  \textbf{\bibinfo{volume}{72}}, \bibinfo{pages}{3439} (\bibinfo{year}{1994}).

\bibitem{mandelstam1991uncertainty}
\bibinfo{author}{Mandelstam, L.} \& \bibinfo{author}{Tamm, I.}
\newblock \bibinfo{title}{The uncertainty relation between energy and time in
  non-relativistic quantum mechanics}.
\newblock In \emph{\bibinfo{booktitle}{Selected papers}},
  \bibinfo{pages}{115--123} (\bibinfo{publisher}{Springer},
  \bibinfo{year}{1991}).

\bibitem{sank2016}
\bibinfo{author}{Sank, D.} \emph{et~al.}
\newblock \bibinfo{title}{Measurement-induced state transitions in a
  superconducting qubit: Beyond the rotating wave approximation}.
\newblock \emph{\bibinfo{journal}{Physical review letters}}
  \textbf{\bibinfo{volume}{117}}, \bibinfo{pages}{190503}
  (\bibinfo{year}{2016}).

\end{thebibliography}
\subsection*{Acknowledgments}
\noindent We thank J.N. Zhang, C.Y. Hsieh, Z.Q. Yin, Z.B. Yang, G.H. Huang, and X. Chen for helpful discussions and comments. We thank the electronics team of Tencent Quantum Lab for preparing the room-temperature electronics.
\subsection*{Author contributions}
\noindent Z.L.Y., S.M.A., and C.Z.L. developed the theory.
S.M.A. and Z.L.Y. performed the experiment. 
All authors contributed to the data analysis and writing of the manuscript. 
\subsection*{Additional information}
\noindent The authors declare no competing financial interests. 
Supplementary information is available for this paper. 
Correspondence and requests for materials should be addressed to S.M.A.(shuomingan@tencent.com).
\newpage

\onecolumngrid
\newpage
\section{Supplementary Note 1. Exact Input-output Theory}\label{SN1}
\begin{figure}[ht]
    \centering
    \includegraphics[width=0.35\textwidth]{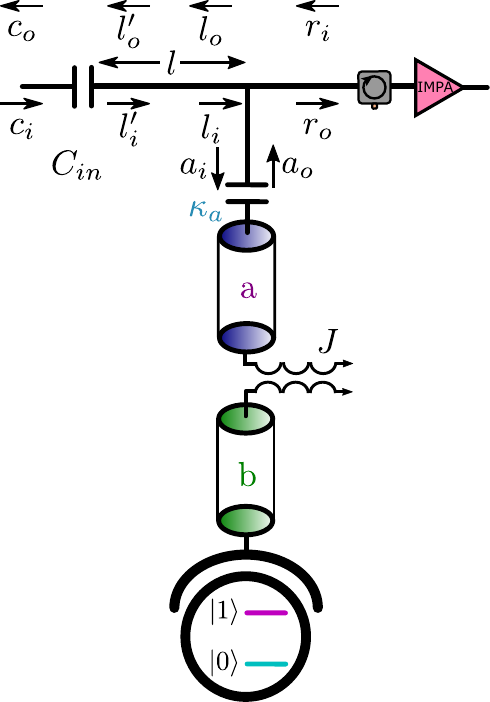}
    \caption{\textbf{Modes in the system.}
    The signal network we analyse here includes the feedline, the filter cavity (mode a), the resonator cavity (mode b) and the qubit. 
    In our feed line, there is an input capacitor to reflect the system leakage to the output port. 
    After the filter, there are three circulators and an impedance modified amplifier. 
    All modes and their directions in the calculation are shown.  
    } 
    \label{fig:SM12_IO}
\end{figure}

Here we derive the exact input-output formula used to simulate the output signal of the system shown in Fig.~\ref{fig:SM12_IO}. Our analysis follows that of Ref.~\cite{heinsoo2018}, although we additionally account for the distance $l$ from the input capacitor $C_{in}$ to the filter $a$, which is necessary for the theory to match experimental observations. 
For wavenumber $k$, the phase accumulated after passing through this distance is $\theta = k\times l$. 
We find that this phase has a profound effect on the final output signal. 
Our goal is to determine how the output field $r_o$ responds to the input field $c_i$ and its interaction with the system, including filter mode $a$, the resonator mode $b$ and the qubit state. To simplify the calculation, instead of directly including the qubit state, we will account for its effect by modifying other system mode frequencies.

To build the mode network, we start from the most left input port and consider the transition and reflection of $C_{in}$ as:
\begin{equation}
    \begin{split}
        l_i' &= (1-\Gamma)c_i +\Gamma l_o'\\
        c_o &= \Gamma c_i + (1-\Gamma) l_o',
    \end{split}
    \label{eqn:Cin}
\end{equation}
where $\Gamma = \frac{Z_l-Z_0}{Z_l+Z_0}$ is the reflection coefficient, $Z_0$ is the impedance of the line and the loaded impedance of $C_{in}$ is $Z_l=\frac{1}{i\omega C_{in}}$. 
As a second step, we consider the effect of the microwave length from $C_{in}$ to the system as:
\begin{equation}
    \begin{split}
        l_i &= e^{i\theta}l_i'\\
        l_o &= e^{-i\theta}l_o'.
    \end{split}
    \label{eqn:l}
\end{equation}
After this, the microwave reaches the T connection between the filter and the feedline. Its scattering matrix is:
\begin{equation}
    \begin{split}
        l_o &= -\frac{1}{3}l_i +\frac{2}{3}r_i+\frac{2}{3}a_o\\
        r_o &= \frac{2}{3}l_i -\frac{1}{3}r_i+\frac{2}{3}a_o\\
        a_i &= \frac{2}{3}l_i +\frac{2}{3}r_i-\frac{1}{3}a_o.
    \end{split}
    \label{eqn:Tmatrix}
\end{equation}
Here we assume the impedance of the line connecting the capacitor is also $Z_0$. Part of the wave in the feedline will drive the filter mode $a$, which satisfies the input-output formula:
\begin{equation}
    a_o = a_i +\sqrt{\kappa_a}a,
    \label{eqn:in-out}
\end{equation}
where $\kappa_a$ is the leakage rate of the mode $a$.
Assuming no output reflection ($r_i = 0$), the relation between the input field $c_i$, the output field $r_o$ and the filter mode $a$ is:
\begin{equation}
    r_o = (1-\Gamma)e^{i\theta}c_i +\frac{1+e^{2i\theta}\Gamma}{2} \sqrt{\kappa_a}a.
    \label{eqn:in-out-a}
\end{equation}
To determine the relation between the input $c_i$ and the output $r_o$ it suffices to deduce how $a$ depends on $c_i$. Under the rotating wave approximation (RWA), the equations of motion in the drive frequency ($\omega_d$) rotating frame are:
\begin{equation}
    \begin{split}
        \dot{a} &= -i\Delta_a a-iJb - \frac{\kappa_a}{2}a-\sqrt{\kappa_a}a_i\\
        \dot{b} &= -i\Delta_b b -iJa,
    \end{split}
    \label{eqn:sys}
\end{equation}
where $\Delta_{a/b} = \omega_{a/b}-\omega_d$ is the detuning of mode $a/b$ frequency $\omega_{a/b}$ relative to the drive frequency $\omega_d$, and $J$ is the coupling strength between modes $a$ and $b$.
It follows from Eq.~\ref{eqn:Cin}, \ref{eqn:l}, \ref{eqn:Tmatrix} and \ref{eqn:in-out} that :
\begin{equation}
    a_i = \frac{1-\Gamma}{2}e^{i\theta}c_i-\frac{(1-e^{i2\theta}\Gamma)}{4}\sqrt{\kappa_a}a.
    \label{eqn:ai-in-a}
\end{equation}
Combining Eq.~\ref{eqn:sys} and \ref{eqn:ai-in-a} gives:
\begin{equation}
    \begin{split}
        \dot{a} &= -i\Delta_a' a-iJb - \frac{\tilde{\kappa}_a}{2}a-\sqrt{\tilde{\kappa}_a}\tilde{a}_i\\
        \dot{b} &= -i\Delta_b b -iJa,
    \end{split}
    \label{eqn:sys-in}
\end{equation}
where the effective detuning, leakage rate, and input field of mode $a$ are $\Delta_a'=\Delta_a+\text{Im}(\Gamma e^{i2\theta})\kappa_a/4$, and $\tilde{\kappa}_a=\kappa_a[1+\text{Re}(\Gamma e^{i2\theta})]/2$, and $\tilde{a}_i=\sqrt{\kappa_a}(1-\Gamma)e^{i\theta}c_i/(2\sqrt{\tilde{\kappa}_a})$ respectively. 
Note that Eq.~\ref{eqn:sys-in} is equivalent, up to redefining various parameters, to Eq.~3 and Eq.~4 of the main text.
According to the designed values of $l$ and $C_{in}$, we estimate $\theta\sim 0.05$ and $\Gamma\sim 0.98-0.17i$.
To simulate the dynamics of modes $a$ and $b$, we use a Lindblad master equation with the Hamiltonian:
\begin{equation}
    H(t) = \Delta_b b^{\dagger}b+\Delta_a' a^{\dagger}a + J(a^\dagger b + b^\dagger a)+i\epsilon^{*}(t)a-i\epsilon(t)a^{\dagger}
    \label{eqn:H}
\end{equation}
and the Lindblad operator $\sqrt{\tilde{\kappa}_a} a$.
The effective driving field $\epsilon(t)$ follows from Eq. \ref{eqn:sys-in} and \ref{eqn:H} as
\begin{equation}
    \epsilon(t) = \sqrt{\tilde{\kappa}_a}\tilde{a}_i,
    \label{eqn:drive}
\end{equation}
which is averaged in the simulation, given the classical (coherent) input field $c_i$.
Substituting this into Eq. \ref{eqn:in-out-a} gives our final input-output formula:
\begin{equation}
    r_o = \frac{2}{\sqrt{\kappa_a}}\epsilon(t)+\frac{1+e^{i2\theta}\Gamma}{2}\sqrt{\kappa_a}a.
    \label{eqn:in-out-drive}
\end{equation}
To account for the weak nonlinearity of the resonator and the uncertainty in the estimated design parameters, in the simulation we multiply the $\sqrt{\kappa_a}a$ term on the right-hand side of Eq.~\ref{eqn:in-out-drive} by a complex coefficient, chosen to fit the simulation to experimental data.

\textbf{Physical interpretation.}
Because the filter mode $a$ driven by $\epsilon(t)$ can be solved using the Lindblad master equation, we can determine the output mode $r_o$ once we know the driving waveform $\epsilon(t)$. 
The physical meaning of Eq.~\ref{eqn:in-out-drive} can be interpreted as follows. 
The factor 2 before $\epsilon(t)$ means only half of the input mode $c'$ is used to drive mode $a$. 
The signal that finally reaches the output port is twice the driving. 
The $e^{i2\theta}\Gamma$ term means half of the leakage of $a$ directly goes to the output port, and the other half will go to the input side and be reflected by $C_{in}$. 
Finally, these two branches interfere with each other and contribute a complex factor between the input and the system leakage.

\section{Supplementary Note 2. Single-Mode Counterdiabatic Driving: Short Derivation \label{SMsec:cd_short}}
Here we give a simple, short derivation of the counterdiabatic (CD) driving (Eq. 1 in the main text) for a single driven bosonic mode coupled to a cold bath, based on a mean-field approximation. We leave a rigorous derivation to Supplementary Note (SN) 3.

As we will see in SN~3, the bosonic mode under consideration can be well approximated by a coherent state, and thus we can use a mean-field approximation for the Heisenberg picture bosonic field $a(t)$, \ie\ $\alpha(t)=\avg{a(t)}$.
Following SN~1, the dynamics are given by the Langevin equation in the drive frame of frequency $\omega_d$:
\begin{equation}
    \dot{\alpha} = -i\Delta_r \alpha - \frac{\kappa}{2}\alpha - \epsilon(t)
    \label{SM2:qle}
\end{equation}
where $\Delta_r\equiv\omega_r-\omega_d$ is the cavity-drive detuning, $\kappa$ is the damping rate due to coupling to the readout line, and $\epsilon(t)$ is the effective drive field. 
The instantaneous equilibrium state is obtained by setting $\dot{\alpha}=0$. 
We denote this instantaneous equilibrium state as:
\begin{equation}
    \bar{\alpha}(t)  = \frac{i\epsilon(t)}{\Delta_r-i\kappa/2},
    \label{SM2:aeq}
\end{equation}
If the drive field is varied slowly enough, the adiabatic theorem guarantees that $\bar{\alpha}(t)$ be the solution of Eq. \ref{SM2:qle}. 
Define the instantaneous diabatic excitation $\delta(t) = \alpha(t)-\bar{\alpha}(t)$.
It follows from Eq. \ref{SM2:qle}, and Eq. \ref{SM2:aeq} and its time derivative, that the dynamics of $\delta(t)$ satisfies:
\begin{equation}
    \dot{\delta}(t)  = -i(\Delta_r-i\kappa/2)\delta(t)- \epsilon_{CD}(t)+\left(\epsilon(t)-i\frac{\dot{\epsilon}(t)}{\Delta_r-i\kappa/2}\right).
    \label{SM2:deq}
\end{equation}
where $\epsilon_{CD}$ is the (new) CD driving. From the boundary conditions $\delta(0)=0$ and $\dot{\delta}(0)=0$, we obtain the desired CD driving as:
\begin{equation}
    \epsilon_{CD}(t) = \epsilon(t) 
    -i \frac{\dot{\epsilon}(t)}{\Delta_r-i\kappa/2}.
    \label{SM2:cddrive}
\end{equation}
Then, for an arbitrarily drive $\epsilon(t)$
, the instantaneous equilibrium state $\bar{\alpha}(t)$ is always the exact dynamic solution of Eq. \ref{SM2:qle}.
\section{Supplementary Note 3. Single-Mode Counterdiabatic Driving: Open Quantum Dynamics Approach}\label{sec:cd_master}
\newcommand{\oper}[2]{\mathcal{#1}(#2)}
\newcommand{\operhat}[2]{\hat{\mathcal{#1}}(#2)}
\newcommand{\opersquare}[2]{\mathcal{#1}[#2]}

\indent
 In this section, we give rigorous derivations of CD driving for a single driven-dissipative bosonic mode,
based on two approaches: (i) Lindblad dynamics~\cite{vacanti2014} and (ii) an adiabatic shortcut of the decoherence free subspace (DFS)~\cite{wu2017adiabatic}. These results justify the mean-field approximation assumed in SN~2, and give additional insight into the adiabatic dynamics of our system.

In what follows, we set $\hbar = 1$. After rotating wave approximation, the Hamiltonian in the driving frame is:
\begin{equation}
    H(t) = \Delta_r a^\dagger a - i\left(\epsilon(t)a^\dagger - \epsilon^*(t) a\right).
\end{equation}
where the cavity-drive detuning $\Delta_r=\omega_r-\omega_d$ depends on the qubit state in the dispersive regime, and $\epsilon(t)$ is the effective drive amplitude.
The transmission line is viewed as a channel for both driving and dissipation, so the dynamics for the cavity density matrix $\rho$ is described by the master equation:
\begin{equation}
    \begin{split}
        \dot{\rho}(t) &= \oper{L}{t}\rho(t)\\
        &= -i[H(t),\rho(t)] + \kappa\opersquare{D}{a}\rho(t)
    \label{SM2:me}
    \end{split}
\end{equation}
where the dissipator is $\opersquare{D}{a}\rho(t) = a\rho(t)a^\dagger - \frac{1}{2}\{\rho(t),a^\dagger a\}$.
Here, only photon decay is considered, since at the effective temperature $T_{mxc} =75 \mK$, the average photon population is $N\approx 0.015 \ll 1$ at readout frequency $\omega_d \approx 2\pi\times6.5 \GHz$. 

\textbf{Lindblad dynamics approach.}
In the adiabatic approximation for open systems~\cite{sarandy2005}, in the limit where the Liouvillian $\mathcal{L}(t)$ is slowly varying, the density matrix $\rho$ evolves independently in each generalized eigenspace of $\mathcal{L}(t)$. 
In other words, $\rho$ can be decomposed into a direct sum of components, one for each independently evolving Jordan block of $\mathcal{L}(t)$. 
The adiabaticity can be made exact by adding a CD Hamiltonian $H_{CD}$ which suppress the inertial part of $\mathcal{L}(t)$ that causes transitions between different Jordan blocks \cite{vacanti2014}.

To determine $H_{CD}$, we first we find a superoperator
$\operhat{O}{t}$ that transforms $\oper{L}{t}$ into Jordan canonical form (JCF). That is, with respect to a certain (not necessarily Hermitian) basis for the density matrix  $B=\{\rho_1,\rho_2,\ldots\}$, we have
\begin{equation}
    \operhat{O}{t}^{-1}\oper{L}{t}\operhat{O}{t} = \operatorname{diag}(J_1(t),J_2(t),\ldots)
\end{equation}
where $J_i(t)$ are the Jordan blocks of size $n_i\times n_i$. 
Second, we transfer to the adiabatic frame defined by $\rho'(t)=\operhat{O}{t}^{-1}\rho(t)$ and show that the non-JCF part of the new Lindblad superoperator $\oper{L'}{t}$, \ie\ $\dot{\rho}'(t)=\oper{L'}{t}\rho'(t)$, can be exactly cancelled by adding a specific CD driving Hamiltonian $H_{CD}(t)$ to the system.

In the first step, we choose $\operhat{O}{t}$ to take the form of a displacement superoperator $\operhat{D}{\alpha(t)}\rho(t) = D(\alpha(t))\rho(t)D(\alpha(t))^{-1}$ where $D(\alpha) \equiv \exp{(\alpha a^{\dagger}-h.c.)}$ is the displacement operator \cite{walls2007}. 
Using the fact that $\operhat{D}{\alpha(t)}a = a - \alpha(t)$, it is straightforward to show that
\begin{align}
     \oper{L_J}{t}\rho(t)&\equiv\operhat{D}{\alpha(t)}^{-1}\oper{L}{t}\operhat{D}{\alpha(t)}\rho(t)\\
     &= -i[H_J(t),\rho(t)]+\kappa\opersquare{D}{a}\rho(t)\\
     H_J(t) &= \Delta_r a^\dagger a + [\Delta_r\alpha(t)-\frac{i\kappa}{2}\alpha(t)-i\epsilon(t)]a^\dagger + h.c.
    \label{SM2:JCF}
\end{align}
Choosing $\alpha(t)=\bar{\alpha}(t)\equiv i\epsilon(t)/(\Delta_r-i\kappa/2)$ -- \ie\ precisely the instantaneous equilibrium state of SN~2, Eq. \ref{SM2:aeq} -- eliminates the time-dependent driving term in $H_J(t)$. Thus, in the adiabatic frame defined by $D(\bar{\alpha}(t))$,
the Liouvillian $\oper{L_J}{t}=\mathcal{L_J}$ is time-independent, so a fixed basis $B$ can be chosen in which $\mathcal{L_J}$ is in JCF.

In the second step, in the adiabatic frame $\rho'(t)=\operhat{D}{\bar{\alpha}(t)}^{-1}\rho(t)$ we have
\begin{equation}
    \begin{split}
    \dot{\rho}'(t)&=-i[i\dot{D}(\bar{\alpha})^{-1}D(\bar{\alpha}),\rho'(t)]+\operhat{D}{\bar{\alpha}}^{-1}\oper{L}{t}\operhat{D}{\bar{\alpha}}\rho'(t)\\
    &= -i[i\dot{D}(\bar{\alpha})^{-1}D(\bar{\alpha}),\rho'(t)]+\mathcal{L_J}\rho'(t).
    \end{split}
    \label{SM2:Ldisplaced}
\end{equation}
\ie\ the dynamics are exactly in JCF except for an inertial Hamiltonian $H_i = i\dot{D}(\bar{\alpha})^{-1}D(\bar{\alpha})$ which mixes the Jordan blocks of $\mathcal{L_J}$. Adding an additional CD term $\mathcal{L}_{CD}\ \rho=-i[H_{CD},\rho]$ to $\mathcal{L}$ exactly cancels $H_i$ if $\operhat{D}{\bar{\alpha}}^{-1}H_{CD}=-H_i$. That is, if
\begin{align}
    H_{CD}(t) &= i\dot{D}(\bar{\alpha}(t))D(\bar{\alpha}(t))^{-1}\\
    &= -\frac{\dot{\epsilon}(t)}{\Delta_r-i\kappa/2}a^\dagger + h.c.
    \label{SM3:cdhamil}
\end{align}
consistent with SN~2, Eq. \ref{SM2:cddrive}.

\textbf{Decoherence free subspace approach.}
The time-dependent decoherence free subspace (DFS) is a subspace of the full system Hilbert space, in which the open system dynamics is unitary and quasi-steady, \ie\ its instaneous motion is generated by an effective Hamiltonian $H_{eff}$ defined within the DFS. By identifying the time-dependent DFS of our system, we derive the CD driving (Eq. \ref{SM3:cdhamil}) and compare it to the Lindblad dynamics approach.

Following the definition in \cite{wu2017adiabatic}, for system dynamics described by a Lindblad master equation $\dot{\rho} = \oper{L}{t}\rho = -i[H(t),\rho] + \sum_{k}\opersquare{D}{a_k(t)}\rho$ where the Lindblad operators $a_k(t)$ have possible time dependence, the time-dependent DFS is the space spanned by a set of orthonormal states $\{\ket{\phi_j(t)}\}$, satisfying: (i) the basis states $\ket{\phi_j(t)}$ are degenerate eigenstates of any Lindblad operator, \ie\ $a_k(t)\ket{\phi_j(t)} = c_k(t)\ket{\phi_j(t)}\ \forall j,k$; (ii) the DFS is closed under the effective Hamiltonian $H_{eff}(t)\equiv H(t) + i/2\sum_{k}c_k^*(t)a_k(t) + h.c.$, \ie\ $H_{eff}$ acting on a state in the DFS results in a state in the DFS. For a single lossy mode described by Eq. \ref{SM2:me}, the DFS exists and is spanned by the single state 
$\ket{\bar{\alpha}(t)}$ for $\bar{\alpha}(t)=i\epsilon(t)/(\Delta_r-i\kappa/2)$, and $H_{eff}$ takes the form of a displaced oscillator $H_{eff}(t)=\Delta_r(a^\dagger-\bar{\alpha}^*(t))(a-\bar{\alpha}(t))$.

Suppose evolution of the DFS is given by the unitary transformation $U(t)$, \ie\ $U(t)\ket{\phi_j(0)} = \ket{\phi_j(t)}$. By direct analogy with closed system CD driving, we can transform to the adiabatic frame defined by $U(t)$ and cancel diabatic excitations out of the DFS by adding a CD Hamiltonian $H_{CD}(t) = i\dot{U}(t)U(t)^\dagger$. In our example, the natural choice for $U(t)$ is the displacement $D(\bar{\alpha}(t))$, from which we can derive the CD Hamiltonian $H_{CD}=i\dot{D}(\bar{\alpha}(t))D(\bar{\alpha}(t))^{-1}$, equivalent to Eq. \ref{SM3:cdhamil} obtained from the Lindblad dynamics approach.

\textbf{Comments.}

\textit{Steady states and adiabatic timescale.} In the adiabatic frame, the Liouvillian $\mathcal{L_J}$ in Eq. \ref{SM2:JCF} takes the form $\mathcal{L_J}\rho= -i[\Delta_r a^\dagger a,\rho]+\kappa\opersquare{D}{a}\rho$. Its eigenvalues can be found by observing
\begin{equation}
    \mathcal{L_J}\ket{m}\bra{n} = [-i\Delta_r(m-n)-\frac{\kappa}{2}(m+n)]\ket{m}\bra{n} + \kappa \sqrt{m n}\ket{m-1}\bra{n-1}
\end{equation}
where $\ket{n}$ are the Fock states, indicating $\mathcal{L_J}$ is upper triangular in the subspace spanned by $\{\ket{0}\bra{j}, \ket{1}\bra{j+1}, \ket{2}\bra{j+2},\ldots\}$ (or their Hermitian conjugates) for natural numbers $j$.
Hence, $\mathcal{L_J}$ has non-degenerate eigenvalues $e_{j,k}=i\Delta_r j-\kappa(j/2+k),\ k=0,1,2,\ldots$ (or their complex conjugates) in each subspace and can be exactly diagonalized. 
We note that the only steady state of $\mathcal{L_J}$,
\ie\ the eigenstate of $\mathcal{L_J}$ with zero eigenvalue, is given by $j=k=0$, which is the vacuum state $\ket{0}$ in the adiabatic frame or the coherent state $\ket{\alpha(t)}$ in the lab frame.

We also comment on the timescale required for the adiabatic approximation to hold in open systems, following the results of~\cite{sarandy2005}. Analogously to closed quantum systems, a sufficient condition for adiabatic evolution of an open system is
\begin{equation}
    \max_{0<t<t_f}|\langle\tilde{\rho}_{j',k'}(t),\dot{\rho}_{j,k}(t)\rangle| \ll |e_{j,k}-e_{j',k'}|,\ \forall j,k,j',k'
\end{equation}
where $t_f$ is the total evolution time, $\langle u,v\rangle\equiv\tr(u^\dagger v)$ defines the inner product, $\rho_{j,k}(t)$ are (lab-frame) eigenstates of $\mathcal{L}(t)$ with eigenvalues $e_{j,k}$, and $\tilde{\rho}_{j,k}(t)$ are eigenstates of $\mathcal{L}(t)^\dagger$. Here the adjoint $\mathcal{L}(t)^\dagger$ is defined as the superoperator that satisfy $\langle \mathcal{L}(t)^\dagger u,v\rangle = \langle u,\mathcal{L}(t) v\rangle,\ \forall u,v$. The LHS is hard to evaluate in practice, and a crude estimate is obtained by setting $\langle\tilde{\rho}_{j',k'}(s),d\rho_{j,k}(s)/ds\rangle\sim 1$ for normalized time $s\equiv t/t_f$.
The adiabatic condition for the total time $t_f$ is then derived as
\begin{equation}
    t_f \gg \frac{1}{\min_{j,k,j',k'}(|e_{j,k}-e_{j',k'}|)} = \frac{1}{\min(\sqrt{\Delta_r^2+\kappa^2/4},\kappa)}.
\end{equation}
For $\Delta_r$ comparable to $\kappa$, which is a usual experimental scenario, the adiabatic condition is $t_f\gg \kappa^{-1}$, making STA useful for fast protocols operating within unit lifetimes. This adiabatic condition is verified in Fig.~\ref{fig:cd_alldurations}, where $\sin^2$-shaped pulses are applied for different durations $t_f$ and $\sin^2$-shaped output signals are observed only for $t_f > 10\kappa^{-1} \approx 600\ns$.
\newline
\newline
\textit{DFS from Lindblad dynamics.}
The derivation of CD driving from both Lindblad dynamics and the DFS approach relies on switching to the adiabatic frame defined by $D(\bar{\alpha}(t))$, \ie\ $\ket{\phi'} = D(\bar{\alpha}(t))^\dagger \ket{\phi}$. We note that the DFS of our system (\ie\ the coherent state $\ket{\bar{\alpha}(t)}$) is the vacuum state $\ket{0}$ in the adiabatic frame - the only steady eigenstate (\ie\ having an eigenvalue with a non-negative real part) of $\mathcal{L}_J$. As a result, the steady eigenspace of the Liouvillian $\mathcal{L}(t)$ is equivalent to the DFS, whereas this is not true in general since purity of these steady states requires a zero-temperature approximation or negligible thermal photon number $N(\omega)\ll 1$ in the frequency band of interest. For bosonic modes in the high temperature regime ($N(\omega)\sim 1$ or $N(\omega)\gg 1$), CD driving is still possible by the Lindblad dynamics approach even though the pure-state DFS does not exist. In this case, although CD driving does not prevent heating into the steady thermal state in the adiabatic frame, it ensures fast transport of this steady state, which is still of practical interest.
\newline
\newline
\textit{Mean Field Approximation.} Here we show that the single driven-dissipative mode remains in a coherent state, which justifies the mean-field approximation used in SN~2. For open quantum systems, the coherent state is known to be the consequence of the zero-temperature approximation of the environment \cite{breuer2002theory}. Specifically, in the frame defined by a general displacement $D(\alpha(t))$, the dynamics in Eq. \ref{SM2:Ldisplaced} can be rewritten as 
    \begin{equation}
        \dot{\rho}' = -i[\Delta_r a^\dagger a + (-i\dot{\alpha}+\Delta_r\alpha-i\frac{\kappa}{2}\alpha-i\epsilon)a^\dagger + h.c.\ , \rho'] + \kappa\opersquare{D}{a}\rho'.
    \label{SM3:full_evolution}
    \end{equation}
Choosing $\alpha(t)$ that satisfies the Langevin dynamics (Eq. \ref{SM2:qle}) thus eliminates the driving term. Consequently, the system stays in the vacuum state in the displaced frame, corresponding to the coherent state $\ket{\alpha(t)}$ in the lab frame.
\section{Supplementary Note 4. Quantum Speed Limit of the CD Driving Protocol}
In this section, we discuss the Quantum Speed Limit (QSL) for a driven-dissipative bosonic mode, and show that our CD driving protocol reaches optimal quantum efficiency among all possible experimental controls. 

For open quantum systems, the QSL can be formulated as a geometric constraint, \ie\ the total length of the system's trajectory is bounded below by the geodesic connecting its initial and final states, where the geometry is defined in terms of the Bures metric \cite{bures1969} for density matrices.
This metric is interpreted as the statistical distinguishability between neighbouring quantum states, expressed in terms of the quantum generalization of Fisher information, \textit{i.e.} the Fisher information maximized over all choices of quantum measurements \cite{braunstein1994statistical}. For our system, the dynamics can be equivalently described by a unitary operator, generated by an effective Hamiltonian $H_{eff}(t)=i\dot{U}(t)U(t)^\dagger$. This reduces the QSL to the Mandelstam-Tamm (MT) bound \cite{mandelstam1991uncertainty}:
\begin{equation}
    \operatorname{arccos}|\langle\phi_i|\phi_f\rangle|
    \leq \int_{t_i}^{t_f} \sqrt{\avg{\Delta H_{eff}^2(t)}} dt
    \label{SM4:MTbound}
\end{equation}
where $\ket{\phi_{i(f)}}$ is the initial(final) state and $\Delta H_{eff} = H_{eff} - \avg{H_{eff}}$. Geometrically, the LHS of Eq. \ref{SM4:MTbound} is the Bures length $s_{Bures}$ of the geodesic joining the initial and final state and the RHS is the integrated total length of the system trajectory whose velocity is given by $ds_{Bures}/dt=\sqrt{F_Q(t)/4}=\sqrt{\avg{\Delta H_{eff}^2(t)}}$ \cite{taddei2013quantum}. Here, $F_Q(t)$ is the quantum Fisher information. We define the quantum efficiency of our protocol to be:
\begin{equation}
    \eta \equiv \frac{\operatorname{arccos}|\langle\phi_i|\phi_f\rangle|}
    {\int_{t_i}^{t_f} \sqrt{\avg{\Delta H_{eff}^2(t)}} dt}
\end{equation}

As shown in Eq. \ref{SM3:full_evolution}, for a general driving $\epsilon(t)$ and the system intialized in the ground state, the dynamics is described by the displacement operator, \ie\  $U(t)=D(\alpha(t))$ where $\alpha(t)$ is the solution to the Langevin equation $\dot{\alpha}(t)=(-i\Delta_r-\kappa/2)\alpha(t)-\epsilon(t)$ (SN 3, Eq. \ref{SM3:full_evolution}). We note that $U(t)$ can generate arbitrary dynamics in the space orthogonal to the system state $\ket{\phi(t)}=\ket{\alpha(t)}$, but the extra freedom can be shown to have no contribution to the uncertainty $\avg{\Delta H_{eff}^2(t)}$. With this choice of $U(t)$ it is straightforward to show that $H_{eff}(t)=i\dot{\alpha}(t)a^\dagger + h.c.$ and $\sqrt{\avg{\Delta H_{eff}^2(t)}}=|\dot{\alpha}(t)|$. 
For CD driving, $\sqrt{\avg{\Delta H_{eff}^2(t)}}$ is simply the added drive $|\epsilon_{CD}(t)-\epsilon(t)|$, which provides the resource for adiabatic speedup in view of the energy-time uncertainty principle. 
Identifying $\ket{\phi_{i,f}}=\ket{\alpha_{i,f}}$ and applying the triangle inequality, we obtain
\begin{equation}
    \eta \leq \frac{\operatorname{arccos}( e^{-|\alpha_f-\alpha_i|^2/2})}
    {|\alpha_f-\alpha_i|}.
    \label{SM4:max_QE}
\end{equation}
\begin{figure}[ht]
    \centering
    \includegraphics[width=0.8\textwidth]{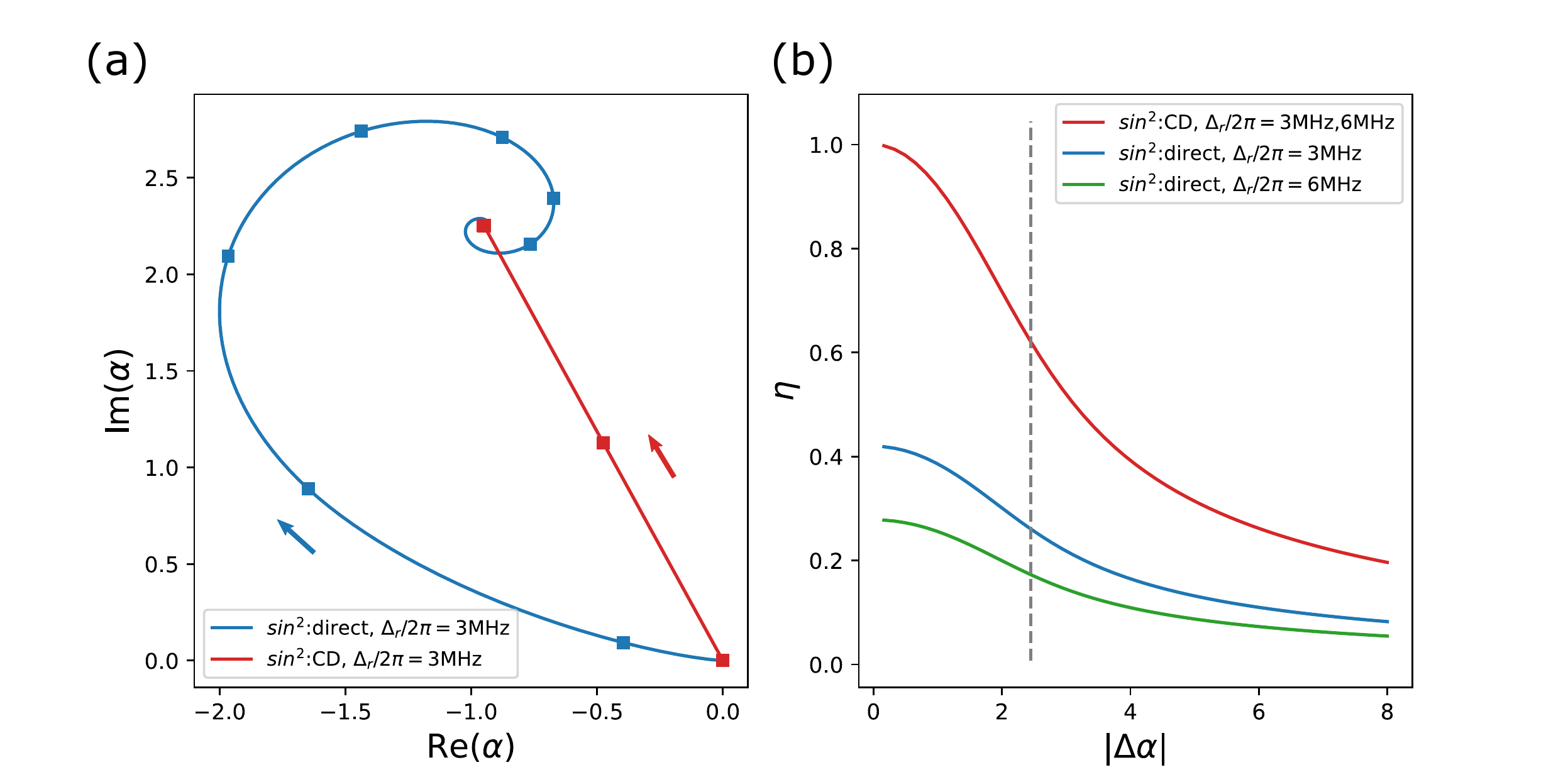}
    \caption{\textbf{Quantum Efficiency of the CD Driving.} (a) Mean field trajectory $\alpha(t)$ for a constant drive preceded by a $t_f=100\ns$ $\sin^2$ ringup, with direct or CD driving. Markers are plotted every 50 ns. CD driving maximizes quantum efficiency by finding the shortest path towards the final state. (b) Quantum efficiency $\eta$ for resonator-drive detunings $\Delta_r/2\pi=3\MHz,\ 6\MHz$, plotted for direct (blue for both detunings) and CD driving (orange for $\Delta_r/2\pi=3\MHz$, green for $\Delta_r/2\pi=6\MHz$). Direct driving with higher detuning is less efficient as it induces spiral trajectories with greater length. The gray line shows the efficiencies for the example in (a).}
    \label{fig:SM4_efficiency}
\end{figure}
with equality achieved by straight-line trajectories -- made possible by CD driving -- in phase space (see Fig.~\ref{fig:SM4_efficiency}~a). The spiral trajectory $\alpha(t)$ in Fig.~\ref{fig:SM4_efficiency}~a is calculated from Eq. \ref{SM2:qle} with parameters $\Delta_r/2\pi=3\MHz$, $\kappa^{-1}=62.88\ns$. Fig.~\ref{fig:SM4_efficiency}~b shows
$\eta$ as a function of target displacement $|\Delta\alpha| = |\alpha_f-\alpha_i|$ for direct and CD driving with two resonator-drive detunings $\Delta_r$. 
For both detunings, the CD driving reaches the optimal quantum efficiency experimentally (as given by the right hand side of Eq.~\ref{SM4:max_QE}). In particular, it saturates the MT bound (\ie\ $\eta\rightarrow1$) in the small driving limit $|\Delta\alpha|\rightarrow 0$. The inefficiency at large $|\Delta\alpha|$ can be explained by the inability to create direct driving to higher-level Fock states, which is a general issue in applying the MT bound for systems with large numbers of energy levels like the bosonic system we consider. Nevertheless, CD driving achieves optimal quantum efficiency within the space of all available pulses, making it favourable for experimental realisation.

\section{Supplementary Note 5. Derivation of the Multi-Mode Optimal Control Protocol}
In this section, we derive the multi-mode optimal control (MMOC) protocol used in the main text, which takes the hybrid frequencies of multiple oscillators as input, and generates a single-port waveform that puts these lossy bosonic modes into thermal equilibrium at a desired final time $t_f$.  We first present a general framework which can be applied to multiple port driving, and then analyse the simpler single port case which is analytically and experimentally more tractable, and sufficient for our needs in the main text.

\textbf{General multiple-port framework.} Consider $n$ linear bosonic modes $\{a_i\}_{i=1}^n$ with frequencies $\omega_i$ and linear couplings $J_{ij}$ between modes $a_i$ and $a_j$. Each mode is coupled to a feedline with strength $\kappa_i$ and driven by an input field $c_i$ at frequency $\omega_d$. In general, the fields $c_i$ can be linearly dependent if they come from the same feedline. In the rotating frame with frequency $\omega_d$ and after rotating wave approximation (RWA), the system Hamiltonian has the form
\begin{equation}
    H_S = \sum_{i=1}^n \Delta_i a_i^\dagger a_i + \sum_{i=1}^n\sum_{j=1,j\neq i}^{n} (J_{ij}a_i^\dagger a_j + h.c.)
\end{equation}
where $\Delta_i=\omega_i-\omega_d$ is the $i$th detuning. In the Heisenberg picture, following the input-output formalism \cite{gardiner1985}, the Langevin dynamics for the $i$th mode is given by $\dot{a}_i=-i[a_i,H_S]-(\kappa_i/2)a_i-\sqrt{\kappa_i}c_i$. Adopting the mean-field approximation $\alpha_i\equiv\avg{a_i}$ for all bosonic modes and defining the effective drive $\epsilon_i\equiv \sqrt{\kappa_i}\avg{c_i}$, we can rewrite the Langevin dynamics in matrix form:
\begin{equation}
    \frac{d}{dt}\Vec{\alpha} +i\Omega\cdot \Vec{\alpha} = -\Vec{\epsilon},
    \ \Omega = 
    \begin{pmatrix}
    \Delta_1-i\kappa_1/2 & J_{12}   & \ldots & J_{1n} \\
    J_{12}^* & \Delta_2-i\kappa_2/2 & \ldots & J_{2n} \\
    \ldots   & \ldots               &        & \ldots \\
    J_{1n}^* & J_{2n}^*             & \ldots & \Delta_n-i\kappa_n/2\\
    \end{pmatrix}
    \label{SM4:InputEOM}
\end{equation}
where $\Omega$ is the (complex) frequency matrix, $\Vec{\alpha}=(\alpha_1,\ldots,\alpha_n)^T$ is the column vector of the mean fields, and $\Vec{\epsilon}(t)=(\epsilon_1(t),\ldots,\epsilon_n(t))^T$ is the column vector of the effective drives. 
$\Omega$ can be diagonalized as $\Omega = O^{-1}\Omega_DO$, where $\Omega_D = \operatorname{diag}(\Delta_1'-i\kappa_1'/2,\ldots,\Delta_n'-i\kappa_n'/2)$ defines the hybrid detunings and linewidths as $\Delta_i'$, $\kappa_i'$.
We use step function driving in our protocol: the time between initial time $t_0$ and final time $t_f=t_m$ is divided into $m$ equal-length intervals, over each of which the drive strength is constant. \ie\ for each drive $\epsilon_i$:
\begin{equation}
    \begin{split}
        \epsilon_i(t<t_0) &= \epsilon_{i0}\\
        \epsilon_i(t_{j-1}<t<t_j) &=\epsilon_{ij},\ \ \ j\in \{1,2,3,\dots, m\}\\
        \epsilon_i(t>t_f) &= \epsilon_{if}.
    \end{split}
    \label{SM4:drive_ansatz}
\end{equation}
where the $\epsilon_{i0}, \epsilon_{ij}, \epsilon_{if}$ are constants. Our goal is to put $\vec{\alpha}(t)$ into the target equilibrium state $\vec{\alpha}_f=i\Omega^{-1}\vec{\epsilon}_f$ at final time $t_f$, starting from initial equilibrium state $\vec{\alpha}_0=i\Omega^{-1}\vec{\epsilon}_0$. The propagator and general solution of differential equation Eq.~\ref{SM4:InputEOM} are 
\begin{align}
    D(t-t')
    &= e^{-i\Omega(t-t')}\theta(t-t') \label{SM4:InputProp}\\
    \vec{\alpha}(t) 
    &= e^{-i\Omega(t-t_0)}\vec{\alpha}(t_0) - \int_{t_0}^t e^{-i\Omega(t-t')}\Vec{\epsilon}(t')dt'\ (t>t_0),
    \label{SM4:GenSol}
\end{align}
where $\theta$ is the step function. Using $\Omega = O^{-1}\Omega_DO$ and Eq. \ref{SM4:GenSol}, our goal can be achieved by solving the equations
\begin{align}
    \sum_{i=1}^n\sum_{j=1}^m O_{ki}G_{kj}\epsilon_{ij} &= \frac{1}{i\Tilde{\Delta}'_k}\sum_{i=1}^n O_{ki}\lp \epsilon_{if}-\epsilon_{i0}e^{-i\Tilde{\Delta}'_k(t_f-t_0)})\rp \label{SM4:constraints}\\
    G_{kj} &\equiv \int_{t_{j-1}}^{t_j}e^{-i\Tilde{\Delta}'_k(t_f-t')}dt',
    \label{SM4:G_matrix}
\end{align}
for $\epsilon_{ij}$, where $\Tilde{\Delta}'_k=\Delta'_k-i\kappa'_k/2$ is the complex hybrid detuning. Treating the piece-wise driving $\epsilon_{ij}$ as a vector $\epsilon_l$ of dimension $n\times m$ and defining the $n\times mn$ matrix $M_{kl}\equiv O_{ki}G_{kj}\ (l=1,2,\ldots,mn)$, Eq. \ref{SM4:constraints} reduces to the linear equations
\begin{align}
    \sum_{l=1}^{mn} M_{kl}\epsilon_l = y_k,\ 
    y_k \equiv \frac{1}{i\Tilde{\Delta}'_k}\sum_{i=1}^n O_{ki}\lp \epsilon_{if}-\epsilon_{i0}e^{-i\Tilde{\Delta}'_k(t_f-t_0)}\rp
    \label{SM4:multiport-constraint}
\end{align}
If complete information of $O$ or $\Omega$ are given, the general solution of Eq. \ref{SM4:multiport-constraint} can be found by performing a singular value decomposition (SVD) of the matrix M. We concentrate instead on the case of single-port driving, which is considerably simpler.

\textbf{Single port driving.} In the special case of single-port driving, all drivings $\epsilon_{i}(t)$ are linearly dependent, and Eq. \ref{SM4:drive_ansatz} reduces to
\begin{equation}
    \begin{split}
        \epsilon_i(t<t_0) &= c_i\epsilon_0\\
        \epsilon_i(t_{j-1}<t<t_j) &=c_i\epsilon_j,\ \ \ j\in \{1,2,3,\dots, m\}\\
        \epsilon_i(t>t_f) &= c_i\epsilon_f.
    \end{split}
    \label{SM4:single_port_ansatz}
\end{equation}
for constant coefficients $c_i$, a single-port driving vector $\epsilon_j$, and boundary conditions $\epsilon_0$, $\epsilon_f$. In this case the $\sum_i O_{ki}c_i$ terms in Eq. \ref{SM4:constraints} cancel, to give 
\begin{align}
    G\cdot\vec{\epsilon} = \vec{y},\ \  
    y_k \equiv \frac{1}{i\Tilde{\Delta}'_k}(\epsilon_f-\epsilon_0e^{-i\Tilde{\Delta}'_k(t_f-t_0)}),
    \label{SM4:reduced_constraints}
\end{align}
which takes the form of a linear constraint on $\vec{\epsilon}$.

Eq.~\ref{SM4:reduced_constraints} can be similarly solved via SVD of the $n\times m$ matrix G, \ie\ $G = U\cdot D\cdot V$ for unitary matrices $U,\ V$ and diagonal matrix $D=(\operatorname{diag}(s_1,\ldots,s_n),0_{n\times (m-n)})$ ($m\geq n$, 0 is the zero matrix). This gives the general form of $\vec{\epsilon}$ as
\begin{equation}
    \vec{\epsilon} = \sum_{i=1}^{n}\frac{(U^{-1}y)_i}{s_i}V^{-1}_i + \sum_{i=n+1}^{m}x_i V^{-1}_i,
    \label{SM4:waveform}
\end{equation}
where $x_{n+1},\ldots,x_{m}$ are free complex parameters and $V^{-1}_i$ is the $i$th column of $V^{-1}$, which can be chosen to optimize a user-defined objective function such as the maximum power output of the pulse (see SN~6). We note that in the single-port driving case, the only input to the protocol is the complex detuning $\tilde{\Delta}_i'$, which is simpler to measure experimentally than the multi-port driving case where full information of $\Omega$ is required.

\textbf{Experimental implementation.}
Our single-port driving experiment in the main text corresponds to 
\begin{equation}
    \vec{a}=(a_0,b_0,a_1,b_1)^T, \vec{\epsilon}=(\epsilon,0,\epsilon,0)^T,\Omega = 
    \begin{pmatrix}
    \Delta_a-i\kappa_a/2 & J              &0   &0\\
    J                    & \Delta_{b,0}   &0   &0\\
    0   &0   &\Delta_a-i\kappa_a/2 & J            \\
    0   &0   &J                    & \Delta_{b,1} \\
    \end{pmatrix}
    \label{SM4:exp_setup}
\end{equation}
where $a_i,b_i$ are the Purcell filter and readout resonator field conditioned on qubit state $i=0,1$, $\Delta_{b,i}$ is the resonator detuning conditioned on qubit state, and $\epsilon$ is the effective driving on the filter port. The drive constants are $c_1=c_3=1,c_2=c_4=0$ in Eq. \ref{SM4:single_port_ansatz}, and the solution to Eq. \ref{SM4:waveform} determines the two quadratures of the driving function which, after optimization over the parameters $x_i$ (discussed in SN~6), yields the waveform used in Fig. 3 of the main text. 

\textbf{Applications.}
Two applications of the class of waveforms derived above are fast equilibration of the readout cavity and Purcell filter and the fast reset of them to the vacuum state. 
In the first case, we set $\epsilon_0=0$ and $\epsilon_f$ in Eq. \ref{SM4:reduced_constraints} to be the constant drive amplitude after $t_f$. 
In the second case, reverse $\epsilon_0$ and $\epsilon_f$.
Unlike the continuous driving pulse in the CD case, the MMOC protocol results in many pulse jumps.
In SN~8, we estimate the effect of the distortion induced by the filter in the AWG and confirm we can still use the MMOC pulses safely.  

\section{Supplementary Note 6. Numerical Optimization and Speed Limit of the MMOC Protocol}
This section covers various numerical aspects of the single-port MMOC protocol of SN~5, including optimization over the maximum power needed, the speed limit of the protocol given limited output power, and the computational complexity of calculating the desired pulse.

\textbf{Energy consumption.}
The total energy consumption (up to an overall constant) of our pulse in Eq. \ref{SM4:waveform} is, due to unitarity of $V^{-1}$,
\begin{equation}
    E(\{x_i\}) 
    \equiv \langle\vec{\epsilon}\ (\{x_i\}),\vec{\epsilon}\ (\{x_i\})\rangle =
    \sum_{i=1}^{n}\left|\frac{(U^{-1}y)_i}{s_i}\right|^2 + \sum_{i=n+1}^{m}|x_i|^2,
    \label{SM5:energy_limit}
\end{equation}
where $\langle\vec{u},\vec{v}\rangle$ denotes the inner product. From Eq.~\ref{SM5:energy_limit} we see that the minimum energy solution $E_{min}$ is obtained by setting $x_i=0$.

\begin{figure}[tbp]
    \centering
    \includegraphics[width=0.4\textwidth]{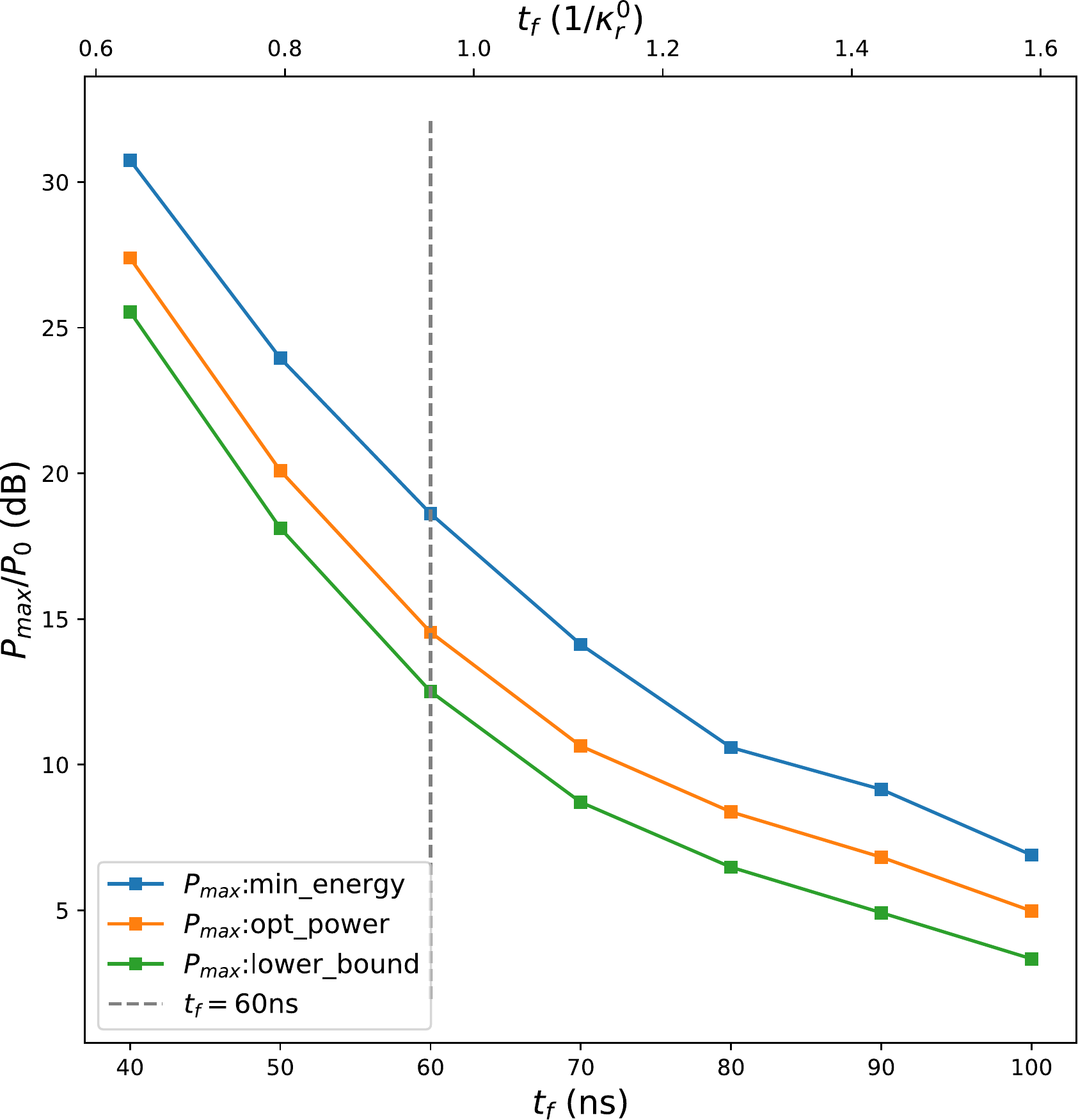}
    \caption{\textbf{Speed-power relation for the MMOC protocol.} Maximum output power $P_{max}$ (in dB) consumed, in units of the steady output power $P_0$ after $t_f$, plotted for different pulses and final times $t_f$. (Blue) The minimum energy pulse, by setting $x_i=0$ in Eq. \ref{SM4:waveform}. (Orange) The pulse whose maximum power is numerically minimized, by exploiting the redundant degrees of freedoms $x_i$. (Green) Theoretical lower bound obtained from Eq. \ref{SM5:power_lb}. The experimental parameters are the same as in the main text.
    }
    \label{fig:SM5_power}
\end{figure}

\textbf{Minimizing the maximum power output.}
Given the output power limitations of the microwave devices, it is desirable to minimize the maximum output power $P_{max}(\{x_i\}) \equiv \operatorname{max}_{i}(|\epsilon_i(\{x_i\})|^2)$ of the pulse. To achieve this, we numerically minimize $P_{max}$ (as a function of free parameters $x_i$ from Eq. \ref{SM4:waveform}) using a differential evolution algorithm. The resulting optimized MMOC pulses for both the ring-up and reset stage are those used in the main text. 

Fig. \ref{fig:SM5_power} shows the numerical results of $P_{max}$ (in dB) in units of the steady power $P_0$ after $t_f$, plotted for the ring-up stage with different protocol times $t_f$. For comparison, we also plot a lower bound on $P_{max}$, which follows from Eq. \ref{SM5:energy_limit} and the fact that $m P_{max}\geq \sum_{i=1}^m |\epsilon_i|^2 \geq E_{min}$:
\begin{equation}
    P_{max}(\{x_i\}) \geq \frac{1}{m}\sum_{i=1}^{n}\left|\frac{(U^{-1}y)_i}{s_i}\right|^2
    \equiv P_{max,lb}
    \label{SM5:power_lb}
\end{equation}
Given the protocol time $t_f=60\ns\approx 1/\kappa_r^0$ used in the main text, we find $P_{max} = 14.5 \dB$ after numerical optimization, which is a $4.1 \dB$ reduction from that of the minimum energy pulse. 
For speedup beyond unit resonator lifetime $1/\kappa_r^0$, $P_{max}$ grows rapidly and may induce unwanted qubit transitions, which sets a speed limit for the MMOC protocol, as discussed in SN~7.

\textbf{Computational complexity.}
For single-port MMOC, the number of total qubit-state-conditioned bosonic modes $n$ (in our experiment $n=4$) is less than the total number $m$ of pulse sections.  In this case, the most time-consuming step in computing Eq.~\ref{SM4:waveform} is the singular value decomposition of the $m\times n$ matrix $G$, which has time complexity $O(mn\min(m,n))=O(mn^2)$. For the general case of $n$-port driving, $G$ is replaced by the $n\times mn$ matrix $M$, with corresponding complexity $O(mn^3)$. In either case, the problem admits an efficient polynomial time solution.

\section{Supplementary Note 7. Influences of the Large Drive}\label{N7}
We observe that the output signal drifts with a large driving power, which sets a limit on the steady-state driving power of our protocols.
This can be explained by the nonlinearity of the resonator \cite{brookes2021critical}.
At the same time, according to the previous study  \cite{sank2016}, higher transmon levels are excited due to the non-RWA part of the qubit-resonator Hamiltonian, which becomes on-resonant as the photon number in the resonator increases through a Raman-like process. 
These two observations are shown to be closely related in theoretical simulations \cite{mavrogordatos2017simultaneous}.
Here, we conduct two different experiments to confirm this point and find limitations of our protocol when applied to the transmon-resonator cQED system.

\begin{figure}[tbp]
    \centering
    \includegraphics[width=0.5\textwidth]{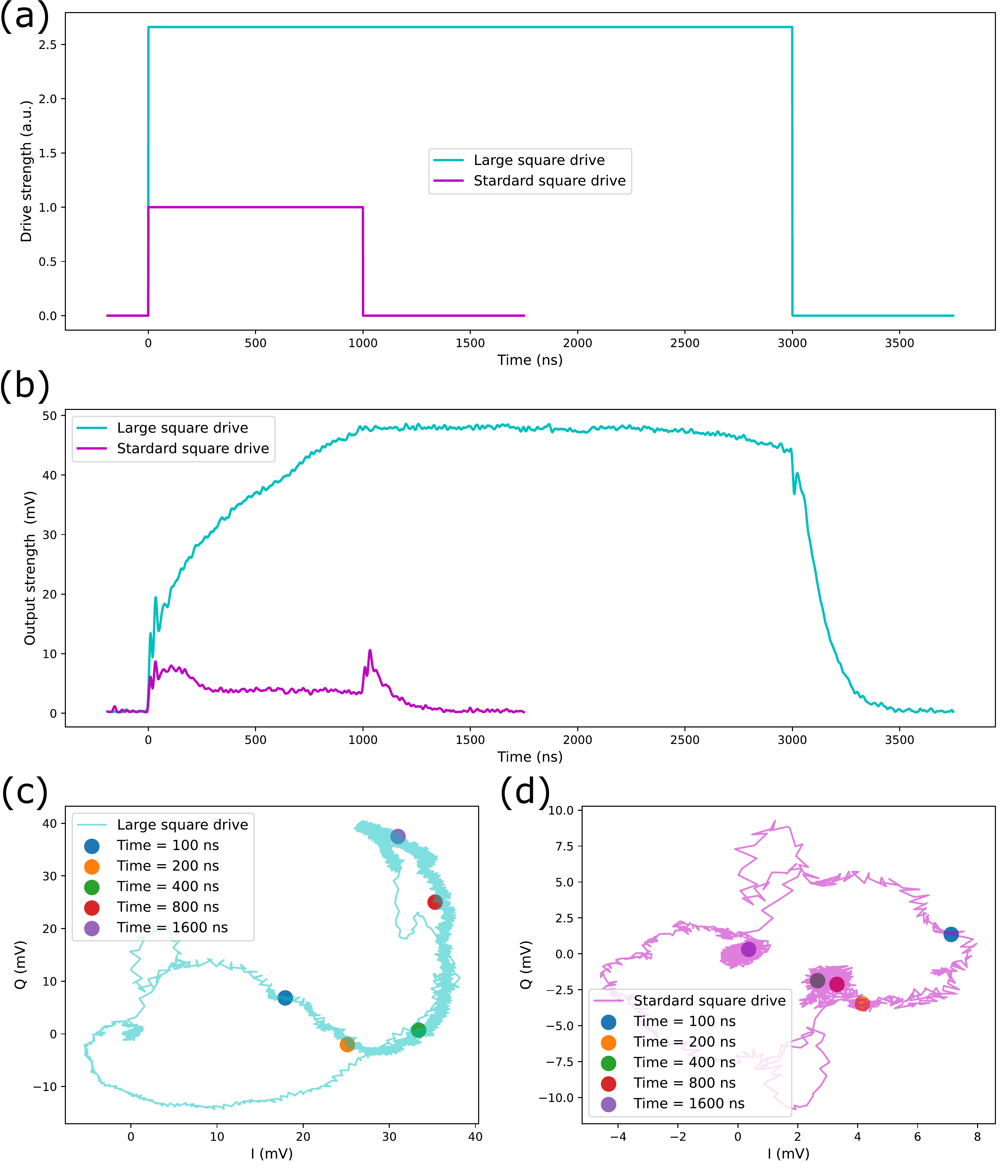}
    \caption{\textbf{Output signal comparison for different pulse amplitudes.} (a) The square-wave drive pulse with large (blue) and small (pink) amplitudes. (b) The average output signal (in mV) for the two pulses, as a function of time. For the small amplitude pulse, steady output is reached after the expected equilibrium time $t_e = 5 \kappa_r^{-1} \approx 300 \ns$. For the large amplitude pulse, the output signal continues to grow well beyond this time. (c,d) The I/Q quadrature trajectories for the (c) large and (d) small amplitude pulse. Various intermediate times are colour-coded in the figure.}
    \label{fig:SM6_IQcomparison}
\end{figure}

\begin{figure}[tbp]
    \centering
    \includegraphics[width=0.6\textwidth]{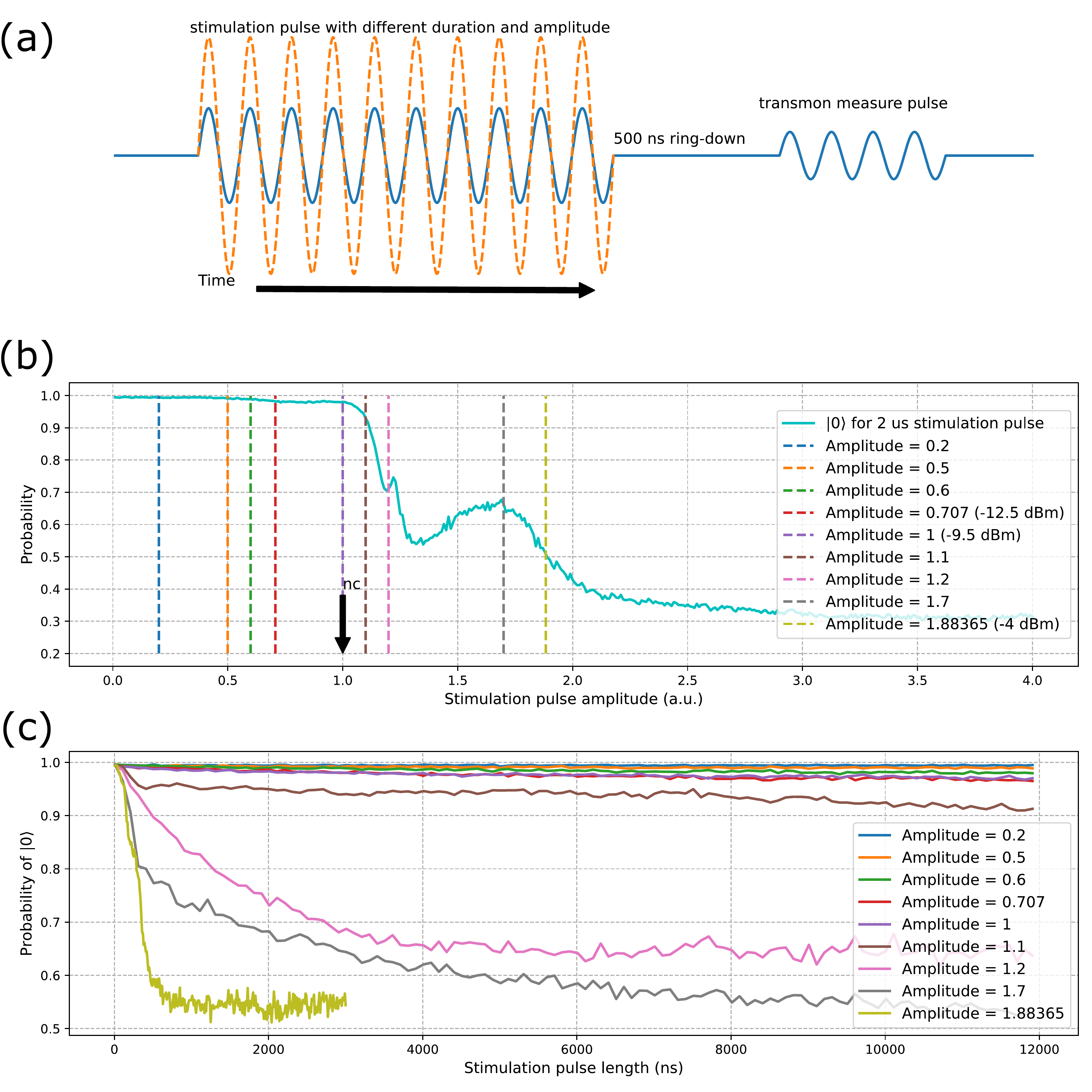}
    \caption{\textbf{Impacts on qubit population induced by resonator excitation.} (a) Pulse sequence for the readout fidelity measurement. A stimulation pulse is first applied with variable strength and duration. After a 500 ns resonator ring-down, a weak measurement pulse is applied to measure the qubit state. (b) Readout fidelity for qubit ground state, $P_0$, as a function of stimulation pulse amplitude. The pulse duration is fixed at $2\ \mu\unit{s}$. A resonance peak (of error) is found between amplitude $1 \unit{a.u.}$ to $1.5 \unit{a.u.}$, similar to the observations of \cite{sank2016}. (c) $P_0$ versus stimulation pulse duration, plotted for various amplitudes marked by coloured lines in (b). The fidelity drops drastically in the first 500 ns when the amplitudes are greater than $1 \unit{a.u.}$}
    \label{fig:SM6_P0impact}
\end{figure}

\begin{figure}[tbp]
    \centering
    \includegraphics[width=0.7\textwidth]{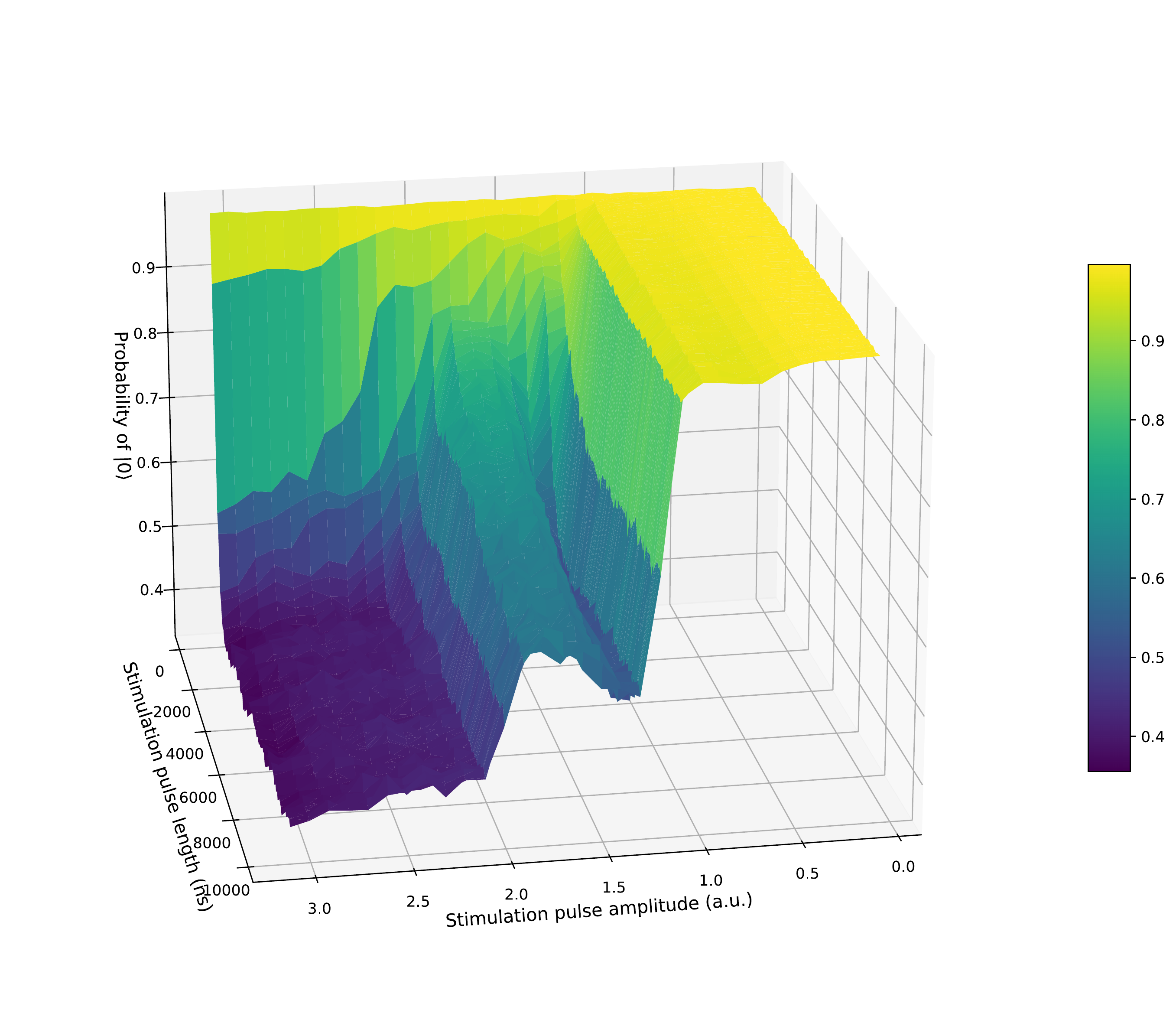}
    \caption{\textbf{Impacts on qubit population induced by cavity excitation, the full data.} Population $P_0$ of the $\left|0\right>$ state, measured as a function of stimulation amplitude and duration, following the procedures in Fig. \ref{fig:SM6_P0impact}. A clear drop in $P_0$ is found for amplitude greater than $1 \unit{a.u.}$.}
    \label{fig:SM6_P0impact3D}
\end{figure}

In the first experiment (Fig.~\ref{fig:SM6_IQcomparison}), we compare the output signal of a small pulse of strength $1$ a.u., and another larger pulse of strength $2.66$ a.u.. 
Each point is averaged over $3\times10^4$ measurements and moving averaged with a Savitzky-Golay filter (width 21, order 3). 
IQ traces of the output signal in Fig.~\ref{fig:SM6_IQcomparison}(c) show a clear drift even long after $5\kappa_a^{-1}$, which can be qualitatively explained by the nonlinearity of the cavity mode.
In Fig.~\ref{fig:SM6_P0impact} and \ref{fig:SM6_P0impact3D}, another experiment is conducted to test the impact on the transition out of the $\left|0\right>$ state of different pulse amplitudes and durations. 
A significant drop in $P_0$ is observed above amplitude $1  \unit{a.u.}$. 
At this drive amplitude we estimate the steady-state cavity photon number (via qubit spectroscopy) to be roughly the critical photon number $n_c\equiv (\Delta/2g)^2\approx 18$ \cite{blais2004cavity}.     

\section{Supplementary Note 8. Propagator Corrections from the
Low-pass Filter in the AWG Driving Line}\label{sec:SN8}
Here we show that corrections to the MMOC pulses imposed by the low-pass fourth-order Chebyshev filter are negligible. In our experiments, the MMOC pulse (Eq.~\ref{SM4:waveform}) from the arbitrary wave generator (AWG) has a carrier driving frequency $\omega_d/(2\pi)$ of $200$ to $250 \MHz$, which passes through the filter with a cutoff frequency $\omega_c/(2\pi) = 750 \MHz$. The piece-wise constant pulse causes the Gibbs phenomenon, a potential source of error. 

Here we give a qualitative evaluation of this error. To simplify our calculations, we assume that the passband's transfer function $g$ is 1, and is 0 outside the passband. The waveform after the filter $\epsilon'(t)$ is described by a convolution $\mathcal{F}$ of the pre-filter waveform $\epsilon(t)$ with the filter function
\begin{equation}
    \epsilon'(t)=\mathcal{F}[\epsilon](t)=\int_{-\infty}^{\infty}\epsilon(\tau) g(t-\tau)d\tau=\int_{-\infty}^{\infty}\epsilon(\tau)d\tau\int_{\omega_0}^{\omega_1}\frac{d\omega}{2\pi}e^{-i\omega(t-\tau)},
\end{equation}
with frequency cutoffs $\omega_0 = \omega_d - \omega_c$ and $\omega_1 = \omega_d + \omega_c$. The calculation is done in the rotating frame defined by $\omega_d$. Replacing $\epsilon$ by $\epsilon'$ in Eq. \ref{SM4:reduced_constraints} results in a modified constraint matrix $G'$, given by $G'\vec{\epsilon}=G\vec{\epsilon}'$, which satisfies
\begin{align}
    G'_{ij}&=\int_{t_0}^{t_1} \mathcal{F}[\theta_j](t) e^{-i\Tilde{\Delta}'_i(t_f-t)}dt\\
    &=\int_{t_{j-1}}^{t_j}e^{-i\Tilde{\Delta}'_i(t_f-\tau)}d\tau
    \int_{\omega_0}^{\omega_1}\frac{d\omega}{2\pi}
    \frac{e^{i(\Tilde{\Delta}'_i-\omega)(t_1-\tau)}
        -e^{i(\Tilde{\Delta}'_i-\omega)(t_0-\tau)}}
        {i(\Tilde{\Delta}'_i-\omega)}\\
    &= e^{-i\Tilde{\Delta}'_i t_f}\int_{\omega_0}^{\omega_1}\frac{d\omega}{2\pi}
    \frac{e^{i\omega t_j}-e^{i\omega t_{j-1}}}{i\omega}
    \frac{e^{i(\Tilde{\Delta}'_i-\omega)t_1}-e^{i(\Tilde{\Delta}'_i-\omega)t_0}}{i(\Tilde{\Delta}'_i-\omega)},
    \label{SM7:final_Gij}
\end{align}
where $[t_0,t_1]$ is the finite time window in which the corrected pulse $\epsilon'$ is integrated over, and $\theta_j(t)$ is the $j$th square function which equals 1 for $t_{j-1}\leq t\leq t_j$ and 0 elsewhere. The single integral Eq.~\ref{SM7:final_Gij} is easy to evaluate numerically and should be compared to the original matrix elements $G_{ij}$. We find that the relative difference between $G'_{ij}$ and $G_{ij}$ is less than $10^{-6}$ and is generally independent of the choice of window time $t_0$, $t_1$. 

\section{Supplementary Note 9. Additional Experimental Data}
\begin{figure}[tbp]
    \centering
    \includegraphics[width=0.7\textwidth]{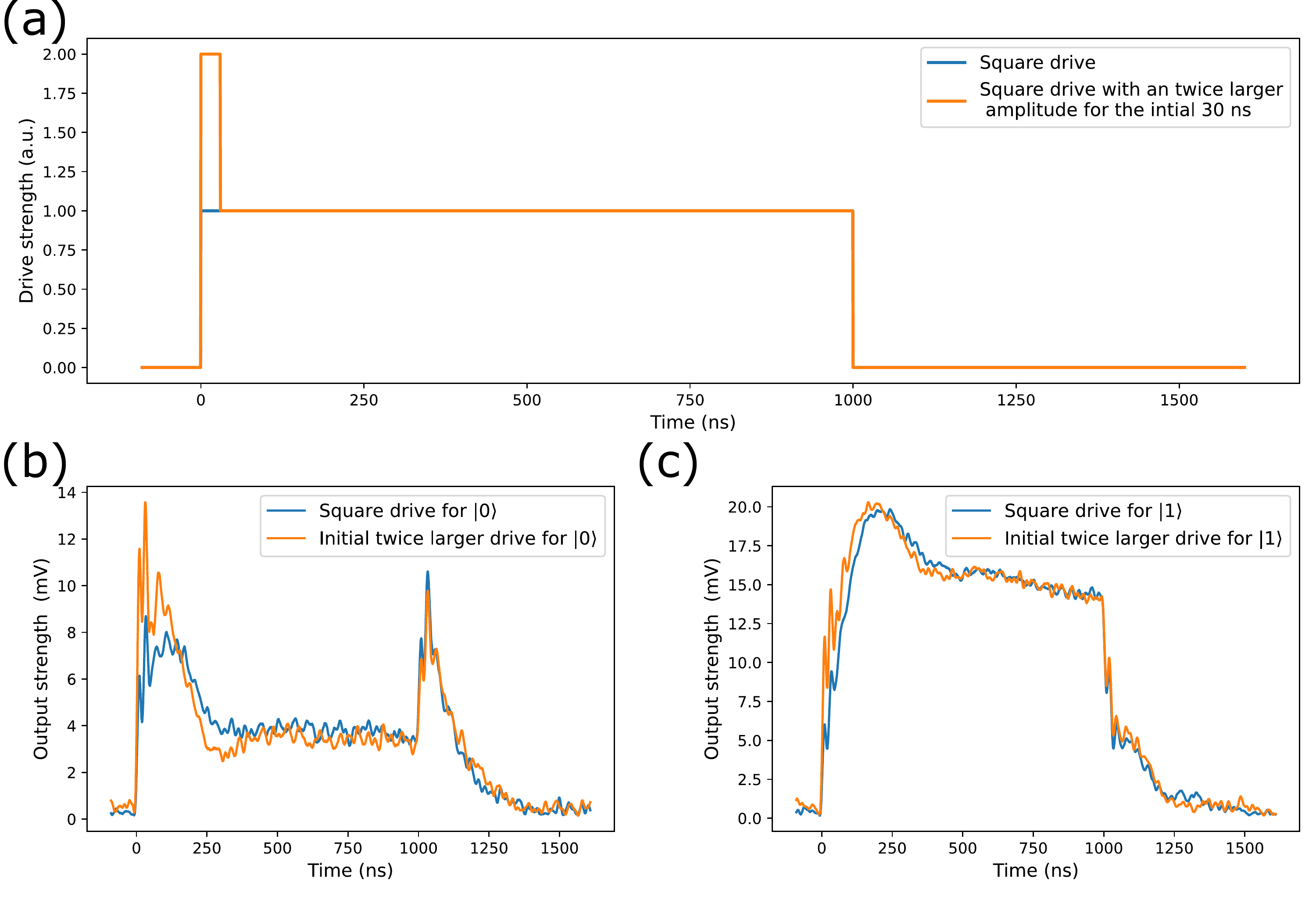}
    \caption{\textbf{Initial larger pulse.} (a) Waveform comparison between the normal driving pulse and a driving pulse with an amplitude $2\times$ larger during the initial 30 ns. In (b) and (c), we see that the initial larger driving pulse does not make the equilibrium process much faster.  
    }
    \label{fig:double_drive}
\end{figure}

\begin{figure}[tbp]
    \centering
    \includegraphics[width=0.8\textwidth]{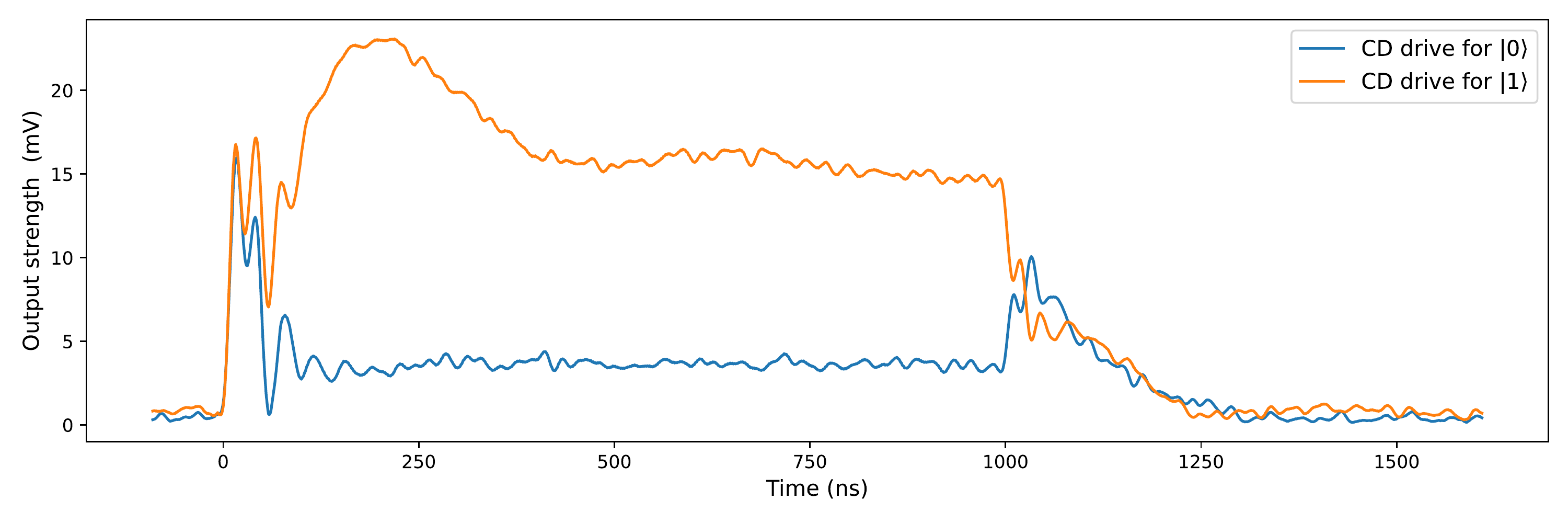}
    \caption{\textbf{Apply the CD driving designed for $\left|0\right>$ on $\left|1\right>$ state.}
    The driving pulse for $\left|0\right>$ is the same as the one used in Fig. 2 in the main text. The same pulse applied when the qubit is prepared in the $\left|1\right>$ state will not accelerate the evolution to equilibrium. 
    }
    \label{fig:cd_0and1}
\end{figure}

\begin{figure}[tbp]
    \centering
    \includegraphics[width=0.6\textwidth]{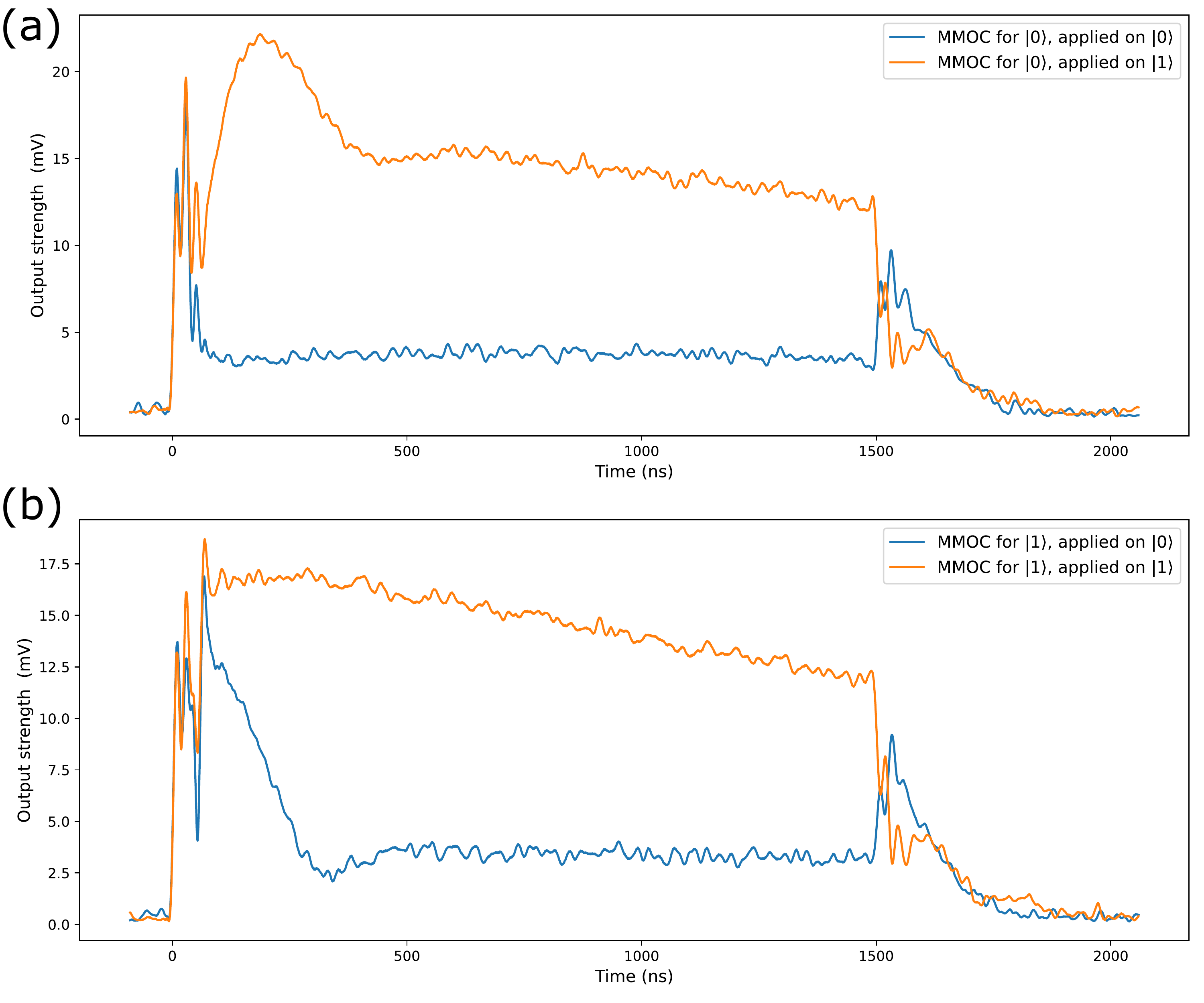}
    \caption{\textbf{Applying the multi-section MMOC designed for $\left|0\right>$($\left|1\right>$) when the qubit is in the $\left|1\right>$($\left|0\right>$) state.}
    In addition to the 4-mode MMOC we used in the main text, we also test the 2-mode MMOC, designed for the hybrid resonator and filter modes with a fixed qubit state.  In (a) and (b), we show the performance of the MMOC for a specific qubit state on both states and see that, as expected, the performance is worse when applied with the qubit in the other state. In (b), the decay of the output from $\approx 100$ for $\left|1\right>$ is caused by the limited lifetime of the qubit.  
    }
    \label{fig:oc_0and1}
\end{figure}

\begin{figure}[tbp]
    \centering
    \includegraphics[width=0.6\textwidth]{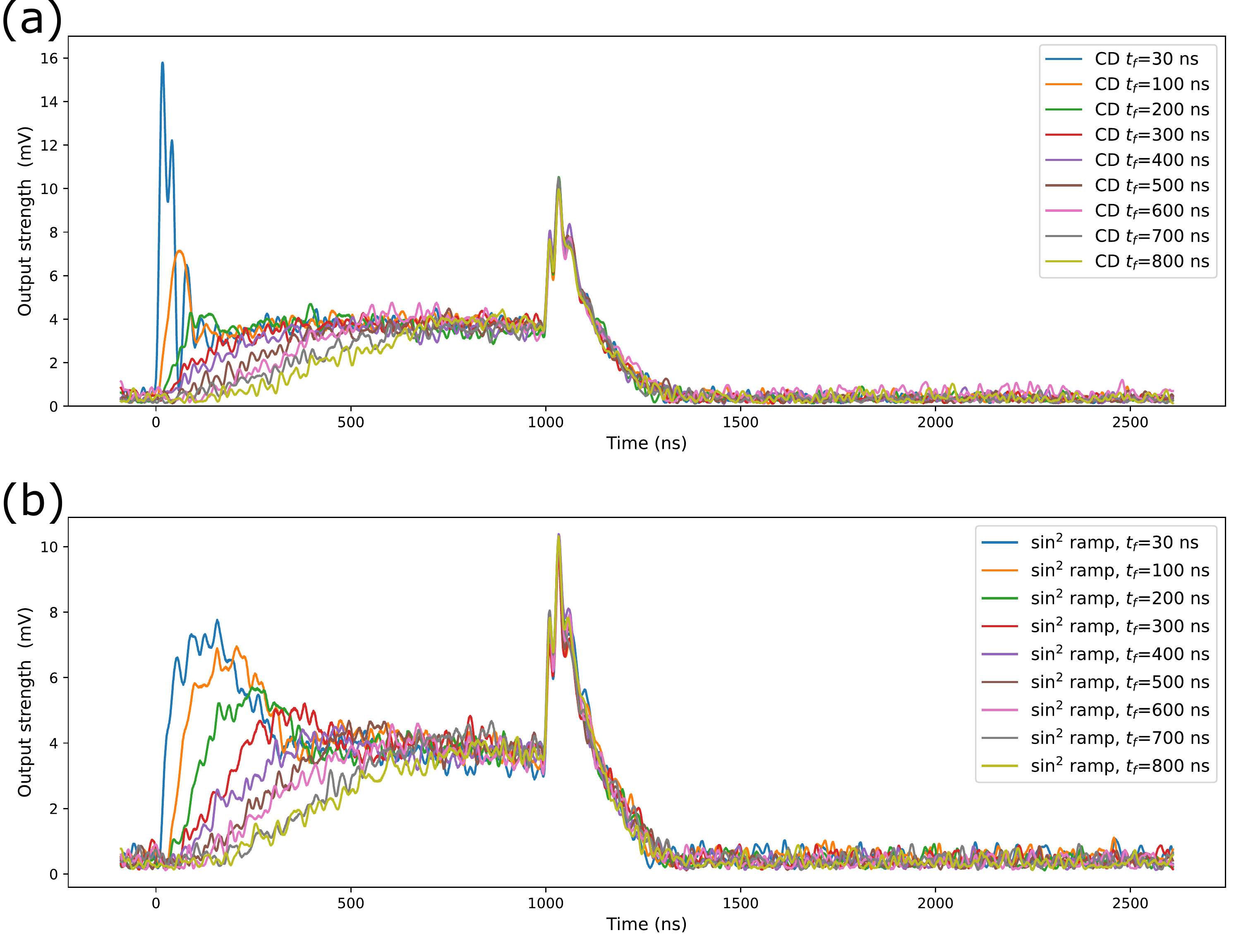}
    \caption{\textbf{CD and $\sin^2$ ringup drives for different durations.}
    Output signal strength for CD and $\sin^2$ pulses with different ring-up durations, using the same parameters and experimental setup as in Fig. 2 of the main text. 
    }
    \label{fig:cd_alldurations}
\end{figure}

\begin{figure}[tbp]
    \centering
    \includegraphics[width=0.7\textwidth]{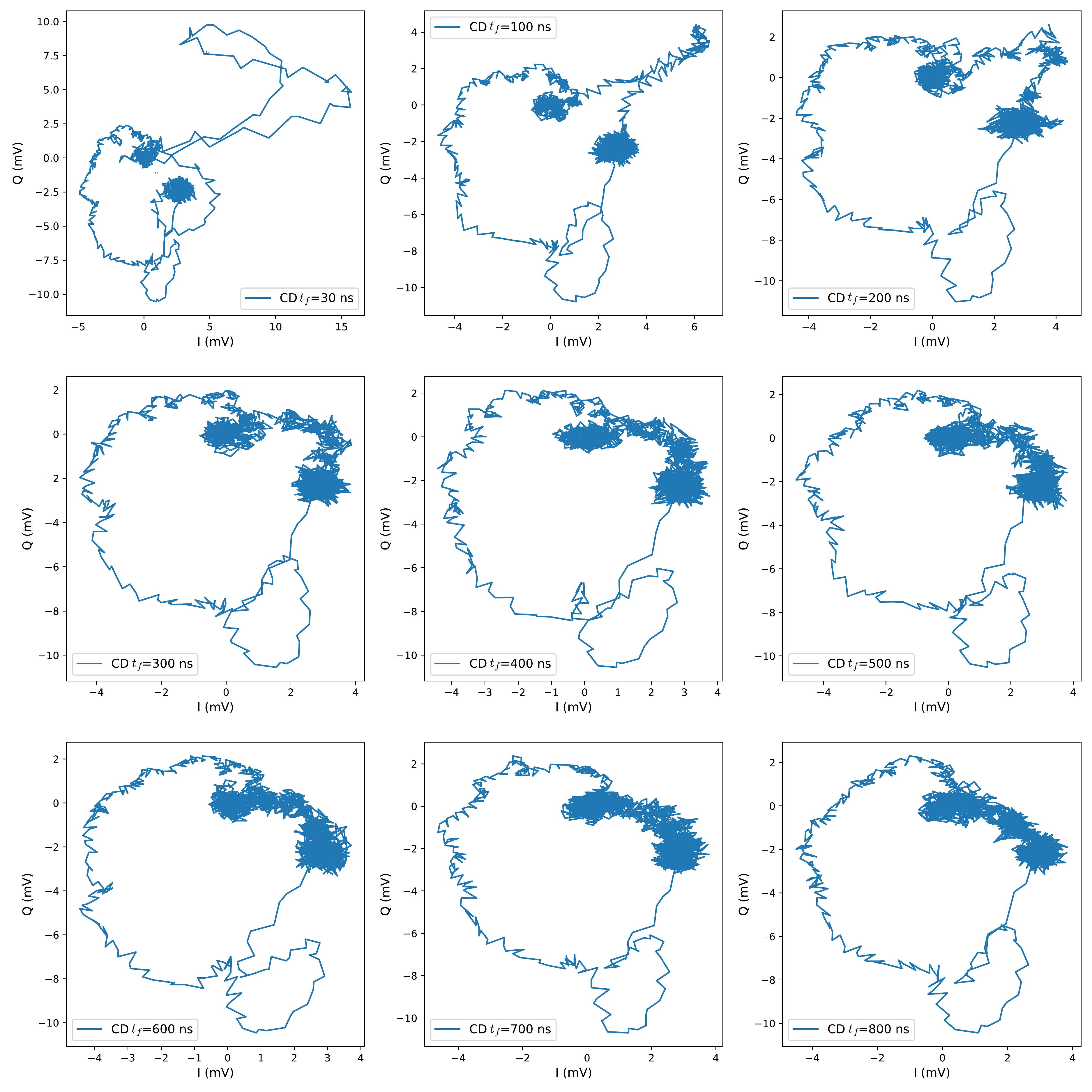}
    \caption{\textbf{I-Q plots of CD driving for different ring-up durations.}
    We see a huge spike during the ringup for a short CD driving, which does not mean the system undergoes highly non-equilibrium dynamics.
    According to the input-output theory, the monitored IQ is the coherent superposition of the input driving and the leakage of the system.
    When the ringup is only $30\ns$, the large amplitude will excite the untargeted filter mode, which is indicated by the spiral collapse to the equilibrium after $30\ns$. 
    }
    \label{fig:cd_IQplots}
\end{figure}

\begin{figure}[ht]
    \centering
    \includegraphics[width=0.7\textwidth]{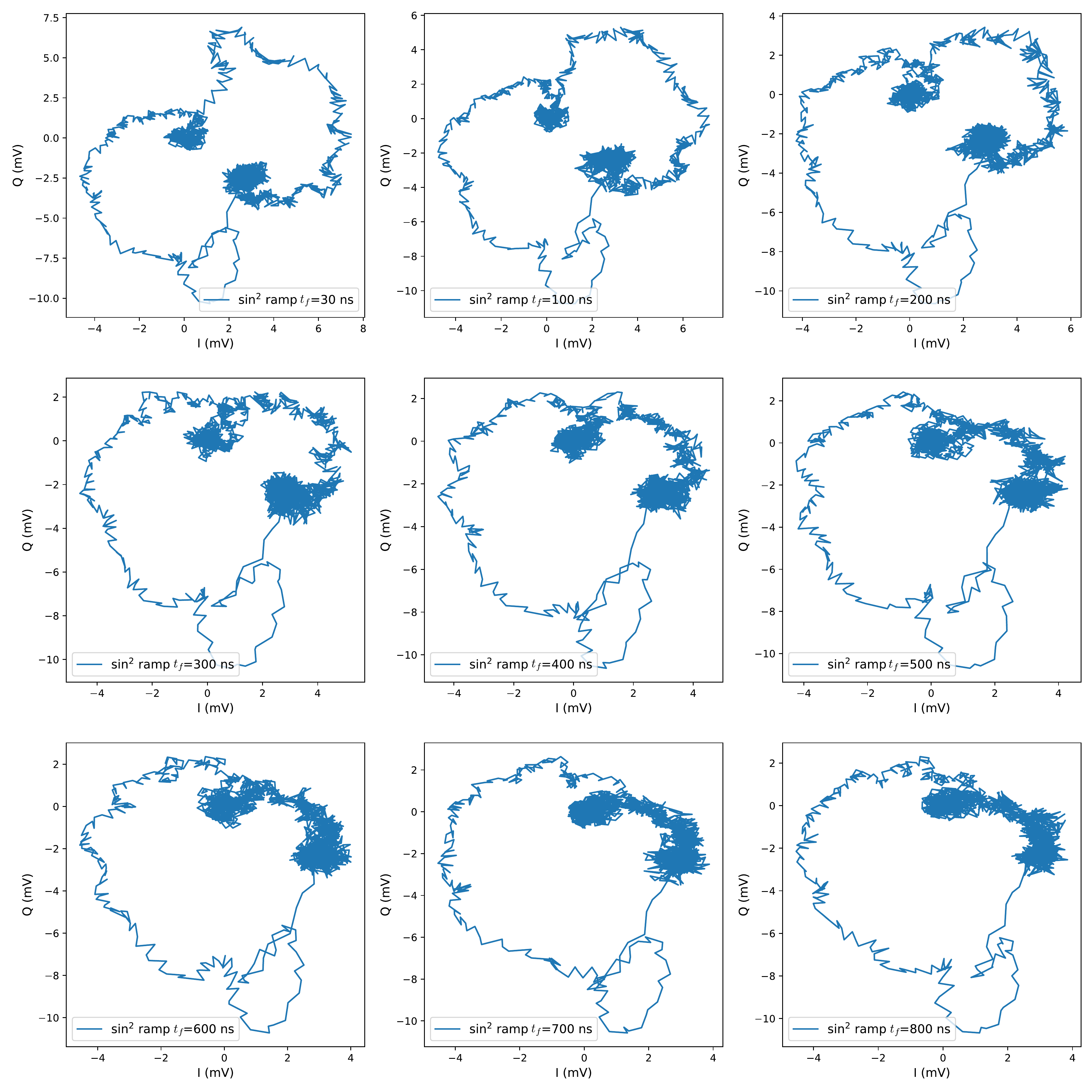}
    \caption{\textbf{I-Q plots of the $\sin^2$ drive for different ringup durations.}
    To see how long it takes to reach quasi-equilibrium dynamics for the corresponding bare $\sin^2$ ringup, we measure the IQ trajectories for different $t_f$. 
    Until $t_f=800\ns$, the trajectory will not change in shape, which indicates the system has reached the quasi-equilibrium dynamics.
    Compared this with the $<100\ns$ in the case of CD driving.
    }
    \label{fig:nocd_IQplots}
\end{figure}

\begin{figure}[ht]
    \centering
    \includegraphics[width=0.7\textwidth]{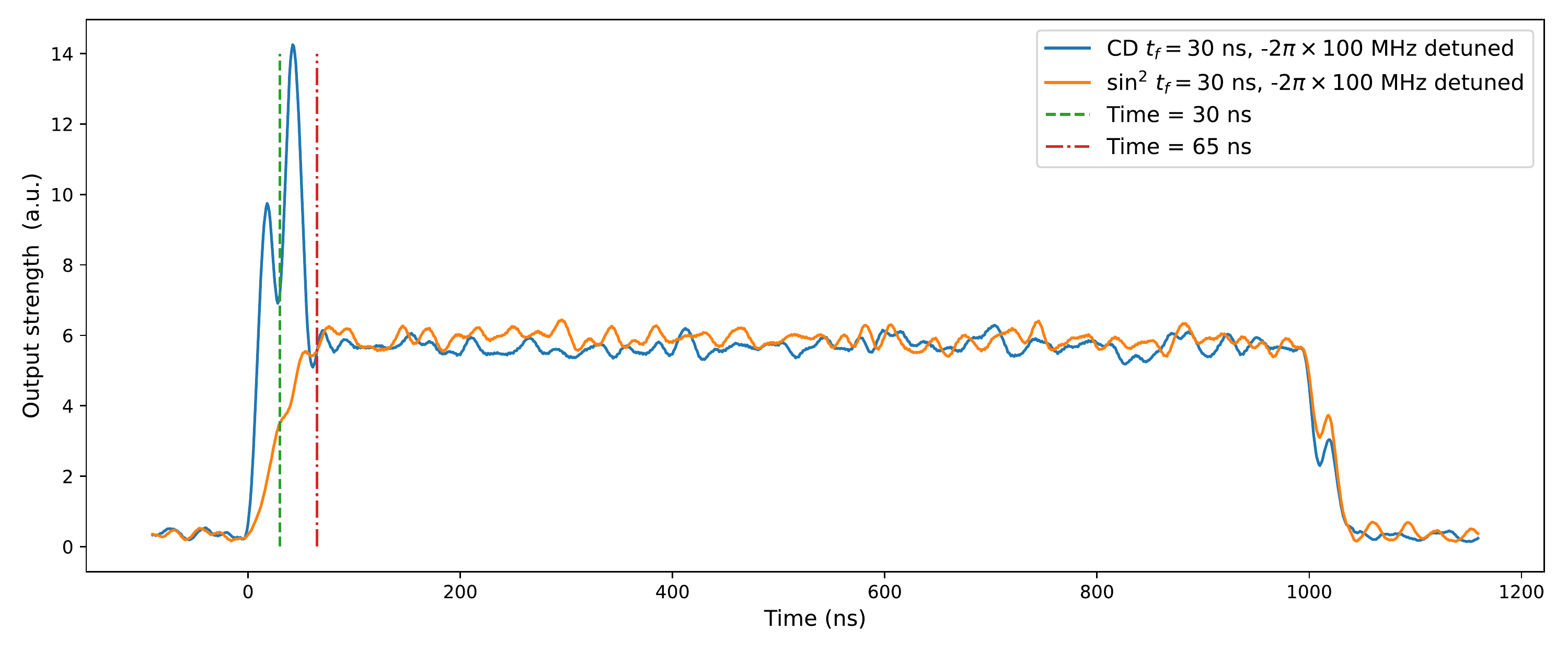}
    \caption{\textbf{Output response of the $-2\pi\times100 \MHz$ detuned CD and $\sin^2$ large-amplitude drivings.}
    When we detune the drive frequency by $-2\pi\times100 \MHz$, the lossy cQED system will not be excited, and the output signal closely follows the input signal. However, we see a delay in the output signal before reaching equilibrium after ringup, and an extended ringdown after the drive has concluded. 
    This is particularly pronounced when the driving power is large.
    }
    \label{fig:squarewave_detuned}
\end{figure}

In this section, we show additional experimental data we have collected. 
In Fig.~\ref{fig:double_drive}, we test the widely used square wave driving with an initial amplitude twice as large as the remaining waveform, and see little decrease in the equilibrium time.
In Fig.~\ref{fig:cd_0and1}, we apply the CD pulse designed according to the set of parameters for the $\left|0\right>$ and $\left|1\right>$ states, and confirm that in the single-port driving situation, CD is only able to accelerate one mode at one time.  
The main text shows the 4-mode MMOC protocol controlling all four modes with the same pulse.
In Fig.~\ref{fig:oc_0and1}, we design the MMOC for two modes corresponding to only one specific qubit state, then apply it on both qubit states.
In this case, the method only works when the qubit is in the correct state.
In Fig.~\ref{fig:cd_alldurations}, we compare the output amplitudes of the $\sin^2$ and the corresponding CD ringup drives of different durations.
Fig.~\ref{fig:cd_IQplots} and Fig.~\ref{fig:nocd_IQplots} show the output signal's IQ trajectories for the $\sin^2$ and the corresponding CD ringup drives.
In Fig.~\ref{fig:squarewave_detuned}, a large-amplitude and far off-resonant $\sin^2$ drive , and corresponding CD drive are applied.
The far-detuning guarantees the cQED system will not be excited.
According to the input-output theory, the output signal will be a simple rescaling of the input signal.
The designed input ringup duration $t_f$ is $30\ns$.
However, we see the output reaches the designed stable value around $t=65\ns$.
This tail indicates the filtering effect of some low-Q and energy-storing microwave components in the feedline.
This effect is more noticeable when the input signal is larger and can be simulated by a convolution with a low-pass filter transfer function. 


\end{document}